\documentclass{article}

\usepackage{amsmath,latexsym,amssymb}
\usepackage[pdftex]{graphicx}
\usepackage{geometry}
\newcommand{\diag}{\ensuremath{\mathrm{diag}}}
\newcommand{\Rank}{\ensuremath{\mathrm{Rank}}\,}
\newcommand{\Ima}{\ensuremath{\mathrm{Im}}\,}
\newcommand{\Ker}{\ensuremath{\mathrm{Ker}}\,}
\newtheorem{theorem}{Theorem}
\newtheorem{example}[theorem]{Example}
\newtheorem{lemma}[theorem]{Lemma}

\begin{document}

\title{\Large{\textbf{A Realization Method for Transfer Functions of Linear Quantum Stochastic Systems Using Static Networks for Input/Output Processing and Feedback
\footnote{This work was funded by the Australian Research Council under grant FL110100020}}}}
\author{Symeon~Grivopoulos \and Ian~Petersen}
\maketitle

\begin{center}
\emph{School of Engineering and Information Technology,}\\
\emph{University of New South Wales at the Australian Defence Force Academy,}\\
\emph{Canberra, ACT 2600, Australia}\\
\textsf{symeon.grivopoulos@gmail.com, i.r.petersen@gmail.com}
\end{center}

\begin{abstract}
The issue of realization of the transfer functions of Linear Quantum Stochastic Systems (LQSSs) is of fundamental importance for the practical applications of such systems, especially as coherent controllers for other quantum systems. So far, most works that addressed this problem have used cascade realizations. In this work, a new method is proposed, where the transfer function of a LQSS is realized by a series of a pre-processing linear static network, a reduced LQSS, and a post-processing linear static network. The introduction of the pre- and post-processing static networks leaves an intermediate reduced LQSS with a simple input/output structure, that is realized by a concatenation of simple cavities. A feedback connection of the cavities through a linear static network is used to produce the correct dynamics for the reduced system. The resulting realization provides a nice structural picture of the system. The key mathematical tool that allows for the construction of this realization, is an SVD-like decomposition for doubled-up matrices in Krein spaces. Illustrative examples are provided for the theory developed.
\end{abstract}

\section{Introduction}
\label{Introduction}

Linear Quantum Stochastic Systems (LQSSs) are a class of models of wide use in quantum optics and elsewhere \cite{garzol00,walmil08,wismil10}. In quantum optics, they describe a variety of devices, such as optical cavities, parametric amplifiers, etc., as well as networks of such devices. The mathematical framework for these models is provided by the theory of quantum Wiener processes and the associated Quantum Stochastic Differential Equations \cite{par99,mey95,hudpar84}. Potential applications of quantum optics include quantum information and photonic signal processing, see e.g. \cite{niechu00,knilafmil01,ral06,zhajam13,zha14}. Another particularly important application of LQSSs is as coherent quantum feedback controllers for other quantum systems, i.e. controllers that do not perform any measurement on the controlled quantum system and thus, have the potential for increased performance compared to classical controllers, see e.g. \cite{yankim03a,yankim03b,jamnurpet08,nurjampet09,maapet11b,mab08,hammab12,critezsoh13}.

A problem of fundamental importance for applications of LQSSs, is the problem of realization/synthesis: Given a LQSS with specified parameters, how does one engineer that system using basic quantum optical devices, such as optical cavities, parametric amplifiers, phase shifters, beam splitters, squeezers etc.? The synthesis problem comes in two varieties. First, there is the \emph{strict realization} problem which we just described. This type of realization is necessary in the case where the states of the quantum system are meaningful to the application at hand. Examples include quantum information processing algorithms \cite{niechu00,knilafmil01,ral06} and state generation \cite{kog12,mawoopetyam14}. In the case that only the input-output relation of the LQSS is important, we have the problem of \emph{transfer function realization}. This is the case, for example, in controller synthesis \cite{mab08,hammab12,critezsoh13}.

In recent years, solutions have been proposed to both the strict and the transfer function realization problems. For the strict problem, \cite{nurjamdoh09,nur10a} propose a cascade of single-mode cavities realization. This allows for arbitrary couplings of the LQSS to its environment. However, not all possible interactions between cavity modes are possible, because the mode of a cavity can influence only modes of subsequent cavities. For this reason, direct Hamiltonian interactions \cite{nurjamdoh09} and feedback \cite{nur10a} between cavities have been used to ``correct'' the dynamics of the cascade to the desired form. For the transfer function realization problem, \cite{pet11,nur10b} have shown that a cascade of single-mode cavities realization is possible for any \emph{passive} LQSS, in which case all cavities needed to realize it are also passive. More recently \cite{nurgripet16}, it has been shown that a cascade of single-mode cavities realization is possible for generic LQSSs.

In this work, we propose an alternative solution to the problem of transfer function realization. We show that by appropriate input and output transformations (which can be realized experimentally by static linear optical networks, see Subsection \ref{Static Linear Optical Devices and Networks}), one needs to realize a much simpler transfer function. This ``reduced'' transfer function can be realized by a concatenation of single-mode cavities in a feedback interconnection through a static linear optical network. In the case of passive LQSSs, this realization is always possible, and all necessary devices needed for it are also passive.

In the case of passive LQSSs, the realization method employs crucially the Singular Value Decomposition theorem of Linear Algebra (SVD) for complex matrices. In order to extend the method to general LQSSs, we prove Theorem \ref{Bogoliubov SVD}, which is an analog of SVD for a class of even-dimensional structured matrices, the so-called doubled-up matrices \cite{goujamnur10,pet10}, in a class of complex spaces with indefinite scalar products, the so-called Krein spaces \cite{gohlanrod83}. The role of the unitary matrices in the SVD, as isometries of the domain and target spaces of the linear map (matrix), is taken up by Bogoliubov matrices (see \cite{goujamnur10} and Subsection \ref{Notation and terminology}) in the case of Krein spaces. This is an example of a new algebraic tool required by the theory of LQSSs (or Linear Quantum Systems Theory), that goes beyond the traditional toolbox of classic Linear Systems Theory. The need for new tools and methods in Quantum Systems Theory is to be expected, since Quantum Systems pose novel challenges compared to classical Systems. Moreover, we expect that Theorem \ref{Bogoliubov SVD}, and especially an equivalent form of it for symplectic spaces, namely Theorem \ref{Symplectic SVD} at the end of Section \ref{Realization of General Linear Quantum Stochastic Systems}, will be of more general mathematical interest.

The rest of the paper is organized as follows: In Section \ref{Background Material}, we establish some notation and terminology used in the paper, and provide a short overview of LQSSs and static linear optical devices and networks. In Section \ref{Realization of Passive Linear Quantum Stochastic Systems}, we demonstrate our method of transfer function realization for passive LQSSs, which is the simplest case. Section \ref{Realization of General Linear Quantum Stochastic Systems} contains the realization result for general LQSSs. Finally, the Appendix contains the proof of Theorem \ref{Bogoliubov SVD}, which is the main technical tool necessary to extend the realization from passive LQSSs to general ones. It also contains some remarks on extensions of the realization method to certain cases not covered by the assumptions of Theorem \ref{Bogoliubov SVD}.

\section{Background Material}
\label{Background Material}

\subsection{Notation and terminology}
\label{Notation and terminology}

We begin by establishing the notation and terminology that will be used throughout this paper:
\begin{enumerate}
  \item $x^{*}$ denotes the complex conjugate of a complex number $x$ or the adjoint of an operator $x$, respectively. As usual, $\Re x$ and $\Im x$ denote the real and imaginary part of a complex number. The commutator of two operators $X$ and $Y$ is defined as $[X,Y]=XY-YX$.
  \item For a matrix $X=[x_{ij}]$ with number or operator entries, $X^{\#}=[x_{ij}^*]$, $X^{\top}=[x_{ji}]$ is the usual transpose, and $X^{\dag}=(X^{\#})^{\top}$. Also, for a vector $x=[x_i]$ with number or operator entries, we shall use the notation $\check{x}=\bigl(\begin{smallmatrix} x \\ x^{\#} \end{smallmatrix}\bigr)$.
  \item The identity matrix in $n$ dimensions will be denoted by $I_n$, and a $r \times s$ matrix of zeros will be denoted by $0_{r \times s}$. $\delta_{ij}$ denotes the Kronecker delta symbol in $n$ dimensions, i.e. $I_n=[\delta_{ij}]$. $\diag(X_1,X_2,\ldots,X_k)$ is the block-diagonal matrix formed by the square matrices $X_1,X_2,\ldots,X_k$. $[Y_1 Y_2 \ldots Y_k]$ is the horizontal concatenation of the matrices $Y_1, Y_2, \ldots, Y_k$ of equal row dimension. $\Ker X$, $\Ima X$, and $\Rank X$ denote, respectively, the kernel (null space), the image (range space), and the rank of a matrix $X$.
  \item We define $J_{2k}=\diag(I_k,-I_k)$, and $\Sigma_{2k} = \bigl(\begin{smallmatrix} 0_{k \times k} & I_k \\ I_k & 0_{k \times k} \end{smallmatrix} \bigr)$. We have that $J_{2k}^2=\Sigma_{2k}^2=I_{2k}$ and, $\Sigma_{2k} J_{2k} \Sigma_{2k}=-J_{2k}$. When the dimensions of $I_n$, $0_{r \times s}$, $J_{2k}$ or $\Sigma_{2k}$ can be inferred from context, they will be denoted simply by $I$, $\mathbf{0}$, $J$ and $\Sigma$.
  \item We define the \emph{Krein space} ($\mathbb{C}^{2k}$, $J_{2k}$) as the vector space $\mathbb{C}^{2k}$ equipped with the \emph{indefinite inner product} defined by $\langle v,w\rangle_J=v^{\dag}J_{2k}w$, for any $v,w \in \mathbb{C}^{2k}$. The $J$-norm of a vector $v \in \mathbb{C}^{2k}$ is defined by $|v|_J = \sqrt{|\langle v,v\rangle_J|}$, and if it is nonzero, a normalized multiple of $v$ is $v/|v|_J$. For a $2r \times 2s$ matrix $X$ considered as a map from ($\mathbb{C}^{2s}$, $J_{2s}$) to ($\mathbb{C}^{2r}$, $J_{2r}$), its adjoint operator will be called $\flat$-\emph{adjoint} and denoted by $X^{\flat}$, to distinguish it from its usual adjoint $X^{\dag}$. One can show that $X^{\flat}=J_{2s}X^{\dag}J_{2r}$. The $\flat$-adjoint satisfies properties similar to the usual adjoint, namely $(x_1 A + x_2 B)^{\flat}=x_1^* A^{\flat} + x_2^* B^{\flat}$, and $(AB)^{\flat}=B^{\flat}  A^{\flat}$.
  \item Given two $r \times s$ matrices $X_1$, and $X_2$, respectively, we can form the $2r \times 2s$ matrix $X=\bigl(\begin{smallmatrix} X_1 & X_2 \\ X_2^{\#} & X_1^{\#} \end{smallmatrix}\bigr)$. Such a matrix is said to be \emph{doubled-up} \cite{goujamnur10}. It is immediate to see that the set of doubled-up matrices is closed under addition, multiplication and taking ($\flat$-) adjoints. Also, $\Sigma_{2r} X \Sigma_{2s}=X^{\#}$, if and only if $X^{2r \times 2s}$ is doubled-up. When referring to a doubled-up matrix $X^{2r \times 2s}$, $X_1^{r \times s}$ and $X_2^{r \times s}$, will denote its upper-left and upper-right blocks.
  \item A $2k \times 2k$ complex matrix $R$ is called \emph{Bogoliubov} if it is doubled-up and $\flat$-unitary, i.e $RR^{\flat}=R^{\flat}R=I_{2m}$. The set of these matrices forms a non-compact Lie group known as the Bogoliubov group. Bogoliubov matrices are isometries of Krein spaces.
\end{enumerate}

\subsection{Linear Quantum Stochastic Systems}
\label{Linear Quantum Stochastic Systems}

The material in this subsection is fairly standard, and our presentation aims mostly at establishing notation and terminology. To this end, we follow the papers \cite{pet10,shapet12}. For the mathematical background necessary for a precise discussion of LQSSs, some standard references are \cite{par99,mey95,hudpar84}, while for a Physics perspective, see \cite{garzol00,garcol85}. The references \cite{nurjamdoh09,edwbel05,goujam09,gougohyan08,goujamnur10} contain a lot of relevant material, as well.

The systems we consider in this work are collections of quantum harmonic oscillators interacting among themselves, as well as with their environment. The $i$-th harmonic oscillator ($i=1,\ldots,n$) is described by its position and momentum variables, $x_i$ and $p_i$, respectively. These are self-adjoint operators satisfying the \emph{Canonical Commutation Relations} (CCRs) $[x_i,x_j]=0$, $[p_i,p_j]=0$, and $[x_i,p_j]=\imath\delta_{ij}$, for $i,j=1,\ldots,n$. We find it more convenient to work with the so-called annihilation and creation operators $a_i=\frac{1}{\sqrt{2}}(x_i + \imath p_i)$, and $a_i^*=\frac{1}{\sqrt{2}}(x_i - \imath p_i)$. They satisfy the CCRs $[a_i,a_j]=0$, $[a_i^*,a_j^*]=0$, and $[a_i,a_j^*]=\delta_{ij}$, $i,j=1,\ldots,n$. In the following, $a=(a_1,a_2,\ldots,a_n)^{\top}$.

The environment is modelled as a collection of bosonic heat reservoirs. The $i$-th heat reservoir ($i=1,\ldots,m$) is described by the bosonic field annihilation and creation operators $\mathcal{A}_i(t)$ and $\mathcal{A}_i^*(t)$, respectively. The field operators are \emph{adapted quantum stochastic processes} with forward differentials $d\mathcal{A}_i(t)= \mathcal{A}_i(t+dt)-\mathcal{A}_i(t)$, and $d\mathcal{A}_i^*(t)= \mathcal{A}_i^*(t+dt)-\mathcal{A}_i^*(t)$. They satisfy the quantum It\^{o} products $d\mathcal{A}_i(t) d\mathcal{A}_j(t)=0$, $d\mathcal{A}_i^*(t) d\mathcal{A}_j^*(t)=0$, $d\mathcal{A}_i^*(t) d\mathcal{A}_j(t)=0$, and $d\mathcal{A}_i(t) d\mathcal{A}_j^*(t)=\delta_{ij} dt$. In the following, $\mathcal{A}=(\mathcal{A}_1,\mathcal{A}_2,\ldots,\mathcal{A}_m)^{\top}$.

To describe the dynamics of the harmonic oscillators and the quantum fields (noises), we need to introduce certain operators. We begin with the class of \emph{annihilator only} LQSSs. We also refer to such systems as \emph{passive} LQSSs, because systems in this class describe optical devices such as damped optical cavities, that do not require an external source of quanta for their operation. First, we have the Hamiltonian operator $H=a^{\dag}Ma$, which specifies the dynamics of the harmonic oscillators in the absence of any environmental influence. $M$ is a $n \times n$ Hermitian matrix referred to as the Hamiltonian matrix. Next, we have the coupling operator $L$ (vector of operators) that specifies the interaction of the harmonic oscillators with the quantum fields. $L$ depends linearly on the annihilation operators, and can be expressed  as $L=Na$. $N$ is called the coupling matrix. Finally, we have the unitary scattering matrix $S^{m \times m}$, that describes the interactions between the quantum fields themselves. In practice, it represents the unitary transformation effected on the heat reservoir modes by a static passive linear optical network that precedes the LQSS, see Subsection \ref{Static Linear Optical Devices and Networks}.

In the \emph{Heisenberg picture} of Quantum Mechanics, the joint evolution of the harmonic oscillators and the quantum fields is described by the following system of \emph{Quantum Stochastic Differential Equations} (QSDEs):
\begin{eqnarray}
da &=& [-\imath M -\frac{1}{2}N^{\dag}N ] \,a\, dt -N^{\dag}S\, d\mathcal{A}, \nonumber \\
d\mathcal{A}_{out} &=& N a\, dt + S\, d\mathcal{A}. \label{Passive LQSS 1}
\end{eqnarray}
The field operators $\mathcal{A}_{i \, out}(t), i=1,\ldots,m$, describe the outputs of the system. We can generalize (\ref{Passive LQSS 1}) by allowing the system inputs to be not just quantum noises, but to contain a ``signal part'', as well. Such is the case when the output of a passive LQSS is fed into another passive LQSS. So we substitute the more general input and output notations $\mathcal{U}$ and $\mathcal{Y}$, for $\mathcal{A}$ and $\mathcal{A}_{out}$, respectively. The forward differentials $d\mathcal{U}$ and $d\mathcal{Y}$ of $m$-dimensional inputs and outputs, respectively, contain quantum noises, as well as linear combinations of variables of other systems. The resulting QSDEs are the following:
\begin{eqnarray}
da &=& [-\imath M -\frac{1}{2}N^{\dag}N ] \,a\, dt -N^{\dag}S\, d\mathcal{U}, \nonumber \\
d\mathcal{Y} &=& N a\, dt + S\, d\mathcal{U}. \label{Passive LQSS 2}
\end{eqnarray}
One can show that the structure of (\ref{Passive LQSS 2}) is preserved under linear transformations of the state $\hat{a}= V a$, if and only if $V$ is unitary. Under such a state transformation, the system parameters $(S,N,M)$ transform according to $(\hat{S},\hat{N},\hat{M})=(S,NV^{-1},VMV^{\dag})$. From the point of view of Quantum Mechanics, $V$ must be unitary so that the new annihilation and creation operators satisfy the correct CCRs.

General LQSSs may contain \emph{active} devices that require an external source of quanta for their operation, such as degenerate parametric amplifiers. In this case, system and field creation operators appear in the QSDEs for system and field annihilation operators, and vice versa. Since these are adjoint operators which have to be treated as separate variables, this leads to the appearance of doubled-up matrices in the corresponding QSDEs. To describe the most general linear dynamics of harmonic oscillators and quantum noises, we introduce generalized versions of the Hamiltonian operator, the coupling operator, and the scattering matrix defined above. We begin with the Hamiltonian operator
\begin{eqnarray*}
H=\frac{1}{2} \left(\begin{array}{l} a \\ a^{\#} \\ \end{array}\right)^{\dag}
\left(\begin{array}{cc} M_1 & M_2 \\ M_2^{\#} & M_1^{\#} \\ \end{array} \right) \left(\begin{array}{l} a \\ a^{\#} \\ \end{array}\right) =\frac{1}{2}\check{a}^{\dag}M\check{a},
\end{eqnarray*}
which specifies the dynamics of the harmonic oscillators in the absence of any environmental influence. The $2n \times 2n$ Hamiltonian matrix $M$ is Hermitian and doubled-up. Next, we have the coupling operator $L$ (vector of operators) that specifies the interaction of the harmonic oscillators with the quantum fields. $L$ depends linearly on the creation and annihilation operators, $L=N_1 a+N_2 a^{\#}$. We construct the doubled-up coupling matrix $N^{2m \times 2n}$ from $N_1^{m \times n}$ and $N_2^{m \times n}$. Finally, we have the Bogoliubov generalized scattering matrix $S^{2m \times 2m}$, that describes the interactions between the quantum fields themselves. In practice, it represents the Bogoliubov transformation effected on the heat reservoir modes by a general static linear quantum optical network that precedes the LQSS, see Subsection \ref{Static Linear Optical Devices and Networks}, and \cite{goujamnur10}.

In the \emph{Heisenberg picture} of Quantum Mechanics, the joint evolution of the harmonic oscillators and the quantum fields is described by the following system of \emph{Quantum Stochastic Differential Equations} (QSDEs):
\begin{eqnarray}
d\check{a} &=& [-\imath JM -\frac{1}{2}N^{\flat}N]\, \check{a} dt - N^{\flat} S d\check{\mathcal{U}}, \nonumber \\
d\check{\mathcal{Y}} &=& N \check{a} dt + S d\check{\mathcal{U}}. \label{General LQSS2}
\end{eqnarray}
The forward differentials $d\mathcal{U}$ and $d\mathcal{Y}$ of $m$-dimensional inputs and outputs, respectively, contain quantum noises, as well as a signal part (linear combinations of variables of other systems). One can show that the structure of (\ref{General LQSS2}) is preserved under linear transformations of the state $\check{\tilde{a}} = V \check{a}$, if and only if $V$ is Bogoliubov. In that case the system parameters $(S,N,M)$ transform according to $(\tilde{S},\tilde{N},\tilde{M})= (S,NV^{-1},(V^{-1})^{\dag}MV^{-1})$. From the point of view of Quantum Mechanics, $V$ must be Bogoliubov so that the new annihilation and creation operators satisfy the correct CCRs.

\begin{figure}[!h]
\begin{center}
\scalebox{.3}{\includegraphics{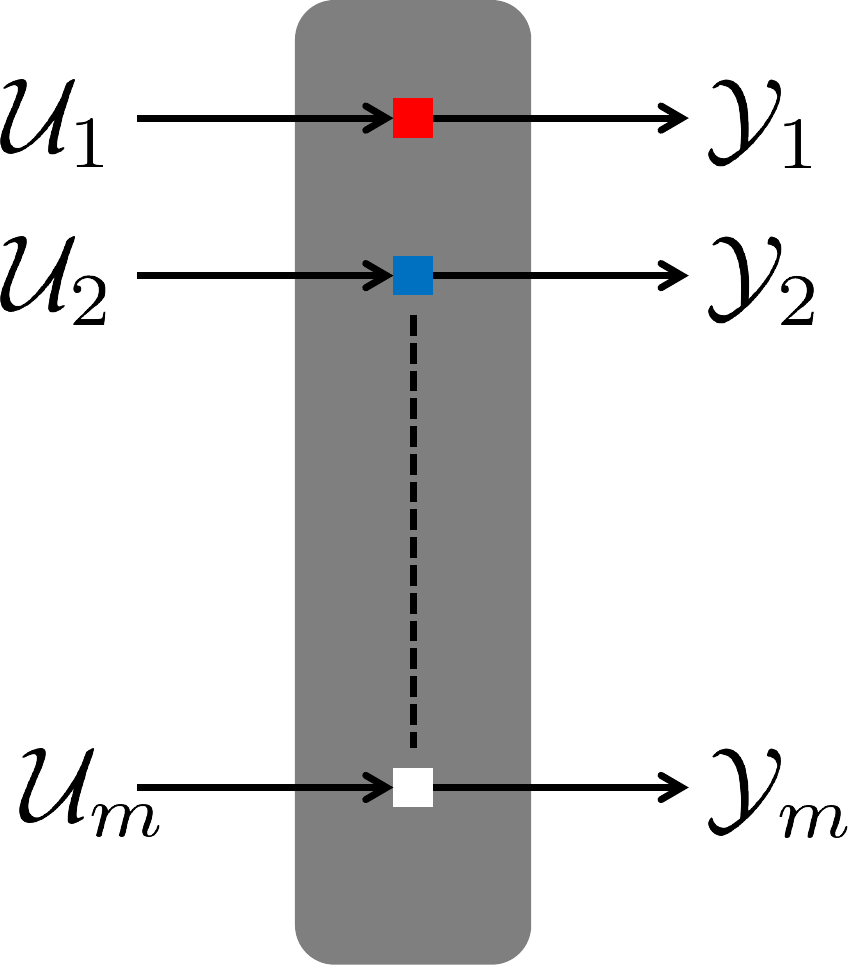}} \caption{Graphical representation of a multi-port cavity. The gray block represents the cavity, and the small squares represent ports. Red is used for passive ports, blue for active ports, and white for all other cases.} \label{General_multiport_cavity}
\end{center}
\end{figure}

We end this subsection with the model of the  single-mode optical cavity, which is the basic device for the proposed realization method in this paper. It is described by its optical mode $a$, with Hamiltonian matrix $M=\diag(\Delta,\Delta)$, where $\Delta \in \mathbb{R}$ is the so-called cavity detuning. For a cavity with $m$ inputs/outputs, we let $N_1=(e^{\imath\phi_1}\, \sqrt{\kappa_1},\ldots,e^{\imath\phi_m}\, \sqrt{\kappa_m})^{\top}$, and $N_2=(e^{\imath\theta_1}\, \sqrt{g_1},\ldots,e^{\imath\theta_m}\, \sqrt{g_m})^{\top}$. $\kappa_i$ and $g_i$ will be called the \emph{passive} and the \emph{active coupling coefficient} of the $i$-th quantum noise to the cavity, respectively. When $g_i=0$, the interaction of the cavity mode with the $i$-th quantum noise will be referred to as \emph{(purely) passive}, and when $\kappa_i=0$, it will be referred to as \emph{(purely) active}. The model of a cavity with $m$ inputs/outputs, is the following:
\begin{eqnarray}
da &=& \Big[-\imath\Delta -\frac{1}{2}\big(N_1^{\dag}N_1 -N_2^{\top}N_2^{\#}\big)\Big] a\, dt \nonumber \\
&-& N_1^{\dag}d\mathcal{U} + N_2^{\top} d\mathcal{U}^{\#} \nonumber \\
&=&\Big( -\imath\Delta -\frac{\gamma}{2} \Big) a\, dt \nonumber \\
&+&\sum_{i=1}^{m} \Big[-e^{-\imath\phi_i}\,\sqrt{\kappa_i}\, d\mathcal{U}_i + e^{\imath\theta_i}\, \sqrt{g_i}\, d\mathcal{U}_i^{\#} \Big], \nonumber \\
d\mathcal{Y}_i &=& e^{\imath\phi_i}\, \sqrt{\kappa_i}\, a \,dt + e^{\imath\theta_i}\, \sqrt{g_i}\, a^{\#} \,dt + d\mathcal{U}_i,  \label{General cavity model}
\end{eqnarray}
$i=1,\ldots,m$, where $\gamma=\sum_{i=1}^{m}(\kappa_i - g_i)$. If a quantum noise couples passively to the cavity, the corresponding interaction may be realized with a partially transmitting  mirror. For an interaction that has an active component, a more complicated implementation is needed, which makes use of an auxiliary cavity, see e.g. \cite{nurjamdoh09} for the details. From now on, we shall use the system-theoretic term \emph{port} for any part of the experimental set-up that realizes an interaction of the cavity mode with a quantum noise (where an input enters and an output exits the cavity). Figure \ref{General_multiport_cavity} is a graphical representation of a  multi-port cavity modelled by equations (\ref{General cavity model}).
%-------------------
%-------------------

\subsection{Static Linear Optical Devices and Networks}
\label{Static Linear Optical Devices and Networks}

Besides the generalized cavities discussed above, our proposed realization method for LQSSs makes use of static linear quantum optical devices and networks, as well. Useful references for this material are  \cite{leo03,nurjamdoh09, leoneu04, bra05}. The most basic such devices are the following:\\[.5em]
1. \textbf{The phase shifter:} This device produces a phase shift in its input optical field. That is, if $\mathcal{U}$ and $\mathcal{Y}$ are its input and output fields, respectively, then $\mathcal{Y}=e^{\imath\theta}\,\mathcal{U}$. Notice that $\mathcal{Y}^* \mathcal{Y}=\mathcal{U}^* \mathcal{U}$. This means that the energy of the output field is equal to that of the input field, and hence energy is conserved. Such a device is called passive.\\[.5em]
2. \textbf{The beam splitter:} This device produces linear combinations of its two input fields. If we denote its inputs by $\mathcal{U}_1$ and $\mathcal{U}_2$, and its outputs by $\mathcal{Y}_1$ and $\mathcal{Y}_2$, then
\[ \left(\begin{array}{c}
\mathcal{Y}_1 \\
\mathcal{Y}_2 \\
\end{array}\right)= R \left(\begin{array}{c}
\mathcal{U}_1 \\
\mathcal{U}_2 \\
\end{array}\right),  \]
where
\[ R=e^{\imath \zeta} \left(\begin{array}{rr}
e^{\imath \frac{\phi + \psi}{2}}\,\cos\frac{\theta}{2} & e^{\imath \frac{\psi-\phi}{2}}\,\sin\frac{\theta}{2} \\[.25em]
-e^{\imath \frac{\phi - \psi}{2}}\,\sin\frac{\theta}{2} & e^{-\imath \frac{\phi + \psi}{2}}\,\cos\frac{\theta}{2} \\
\end{array}\right). \]
$0\leq\theta<2\pi$ is called the mixing angle of the beam splitter. $\phi$ and $\psi$ are phase differences in the two input and the two output fields, respectively, produced by phase shifters. $\zeta$ is a common phase shift in both output fields. This form of $R$ corresponds to a general $2 \times 2$ unitary matrix. Because of this, we can see that
\begin{eqnarray*}
\left(\begin{array}{cc}
\mathcal{Y}_1^* & \mathcal{Y}_2^* \\
\end{array}\right)\left(\begin{array}{c}
\mathcal{Y}_1 \\
\mathcal{Y}_2 \\
\end{array}\right)=
\left(\begin{array}{cc}
\mathcal{U}_1^* & \mathcal{U}_2^* \\
\end{array}\right) R^{\dag}R \left(\begin{array}{c}
\mathcal{U}_1 \\
\mathcal{U}_2 \\
\end{array}\right)=
\left(\begin{array}{cc}
\mathcal{U}_1^* & \mathcal{U}_2^* \\
\end{array}\right)\left(\begin{array}{c}
\mathcal{U}_1 \\
\mathcal{U}_2 \\
\end{array}\right),
\end{eqnarray*}
and hence the total energy of the output fields is equal to that of the input fields. So, the beam splitter is also a passive device.\\[.5em]
3. \textbf{The squeezer:} This device reduces the variance in the real quadrature $\mathcal{(U+U^*)}/2$, or the imaginary quadrature $\mathcal{(U-U^*)}/2\imath$ of an input field $\mathcal{U}$, while increasing the variance in the other. Its operation is described by
\[ \left(\begin{array}{c}
\mathcal{Y} \\
\mathcal{Y}^* \\
\end{array}\right)= R \left(\begin{array}{c}
\mathcal{U} \\
\mathcal{U}^* \\
\end{array}\right),  \]
where
\[ R=\left(\begin{array}{cc}
e^{\imath (\phi + \psi)}\,\cosh x & e^{\imath (\psi - \phi)}\,\sinh x \\[.25em]
e^{\imath (\phi - \psi)}\,\sinh x & e^{-\imath (\phi + \psi)}\,\cosh x \\\end{array}\right). \]
$x \in \mathbb{R}$ is the squeezing parameter, and $\phi,\psi$ are phase shifts in the input and the output field, respectively, produced by phase shifters. This form of $R$ corresponds to a general $2 \times 2$ Bogoliubov matrix. We compute
\begin{eqnarray*}
\mathcal{Y}^* \mathcal{Y}&=&
\frac{1}{2}\left(\begin{array}{cc}
\mathcal{Y}^* & \mathcal{Y} \\
\end{array}\right)\left(\begin{array}{c}
\mathcal{Y} \\
\mathcal{Y}^* \\
\end{array}\right) \\
&=&\frac{1}{2}
\left(\begin{array}{cc}
e^{-\imath\phi}\mathcal{U}^* & e^{\imath\phi}\mathcal{U} \\
\end{array}\right) \left(\begin{array}{cc}
\cosh 2x & \sinh 2x \\
\sinh 2x & \cosh 2x \\\end{array}\right)
\left(\begin{array}{c}
e^{\imath\phi}\mathcal{U} \\
e^{-\imath\phi}\mathcal{U}^* \\
\end{array}\right)\\
&\neq& \frac{1}{2}
\left(\begin{array}{cc}
e^{-\imath\phi}\mathcal{U}^* & e^{\imath\phi}\mathcal{U} \\
\end{array}\right) \left(\begin{array}{c}
e^{\imath\phi}\mathcal{U} \\
e^{-\imath\phi}\mathcal{U}^* \\
\end{array}\right)=\mathcal{U}^* \mathcal{U},
\end{eqnarray*}
for $x\neq 0$, and hence energy is not conserved. So, the squeezer is an active device.

\begin{figure}[!h]
\begin{center}
\scalebox{.3}{\includegraphics{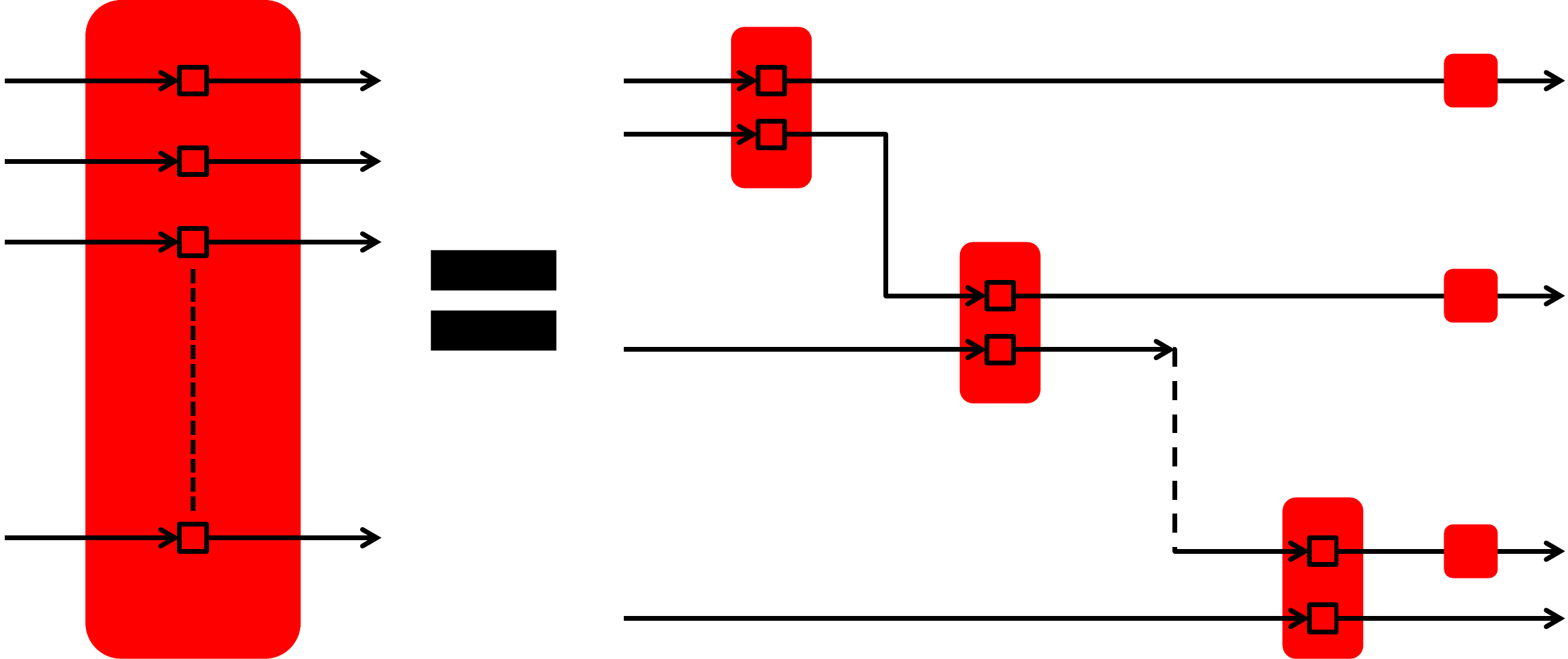}} \caption{Graphical representation of a passive network and its decomposition in terms of beam splitters and phase shifters. In subsequent figures, red blocks will always represent passive static devices and networks.} \label{Passive_network_decomposition}
\end{center}
\end{figure}

\begin{figure}[!h]
\begin{center}
\scalebox{.3}{\includegraphics{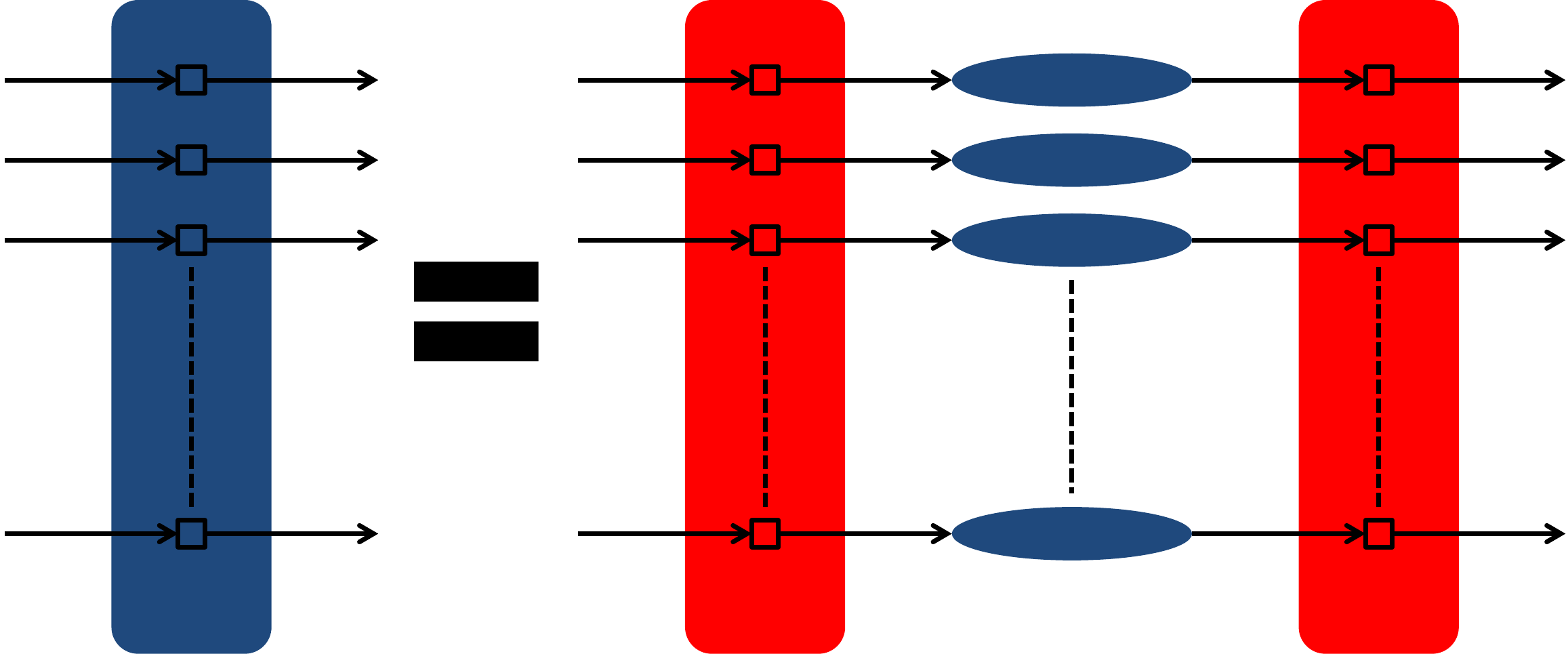}} \caption{Graphical representation of a general network and its decomposition in terms of passive networks and squeezers. In subsequent figures, blue blocks will always represent active static devices and networks.} \label{General_network_decomposition}
\end{center}
\end{figure}

By connecting various static linear optical devices, we may form static linear optical networks (multi-port devices). For a static linear optical network, the relation between the $m$ inputs $\mathcal{U}= (\mathcal{U}_1,\ldots,\mathcal{U}_m)^T$, and the $m$ outputs $\mathcal{Y}=(\mathcal{Y}_1,\ldots,\mathcal{Y}_m)^T$ is linear. For a passive network, i.e. one composed solely of passive devices, we have that $\mathcal{Y}=R\mathcal{U}$, where $R$ is a $m \times m$ unitary matrix. Such a network is a multi-dimensional generalization of the beam splitter and is sometimes called a multi-beam splitter. It turns out that any passive static network can be constructed exclusively from phase shifters and beam splitters \cite{reczeiber94}. This is due to the fact that an $m \times m$ unitary matrix can be factorized in terms of matrices representing either phase shifting of an optical field in the network or beam splitting between two optical fields in the network, see Figure \ref{Passive_network_decomposition}.
%-------------------
%-------------------
In the general case, where the network may contain active devices as well, we have the more general relation $\check{\mathcal{Y}}= R \check{\mathcal{U}}$, where $R$ is a $2m \times 2m$ Bogoliubov matrix. For every Bogoliubov matrix, the following factorization holds:
\[ R= \left(\begin{array}{cc}
U_2 & \mathbf{0} \\
\mathbf{0} & U_2^{\#} \\
\end{array}\right) \left(\begin{array}{cc}
\cosh X & \sinh X \\
\sinh X & \cosh X \\
\end{array}\right) \left(\begin{array}{cc}
U_1 & \mathbf{0} \\
\mathbf{0} & U_1^{\#} \\
\end{array}\right), \]
where $U_1, U_2$ are $m \times m$ unitary and $X=\diag(x_1,x_2,\ldots,x_m)$, with real $x_i, i=1,\ldots,m$. This factorization is known as Bloch-Messiah reduction \cite{nurjamdoh09,leoneu04,bra05}.  The physical interpretation of this equation is that any general static network may be implemented as a sequence of three static networks: First comes a passive static network (multi-beam splitter) implementing the unitary transformation $U_1$. Then follows an active static network made of $m$ squeezers, each acting on an output of the first network, and finally, the outputs of the squeezers are fed into a second multi-beam splitter implementing the unitary transformation $U_2$. This is depicted in Figure \ref{General_network_decomposition}. Because of this structure, a general static network is sometimes called a multi-squeezer. We should stress that both factorizations depicted in Figures \ref{Passive_network_decomposition} and \ref{General_network_decomposition} are constructive, hence arbitrary static linear networks can be synthesized.
%-------------------
%-------------------

\section{Realization of Passive Linear Quantum Stochastic Systems}
\label{Realization of Passive Linear Quantum Stochastic Systems}

\begin{figure}[!h]
\begin{center}
\scalebox{.3}{\includegraphics{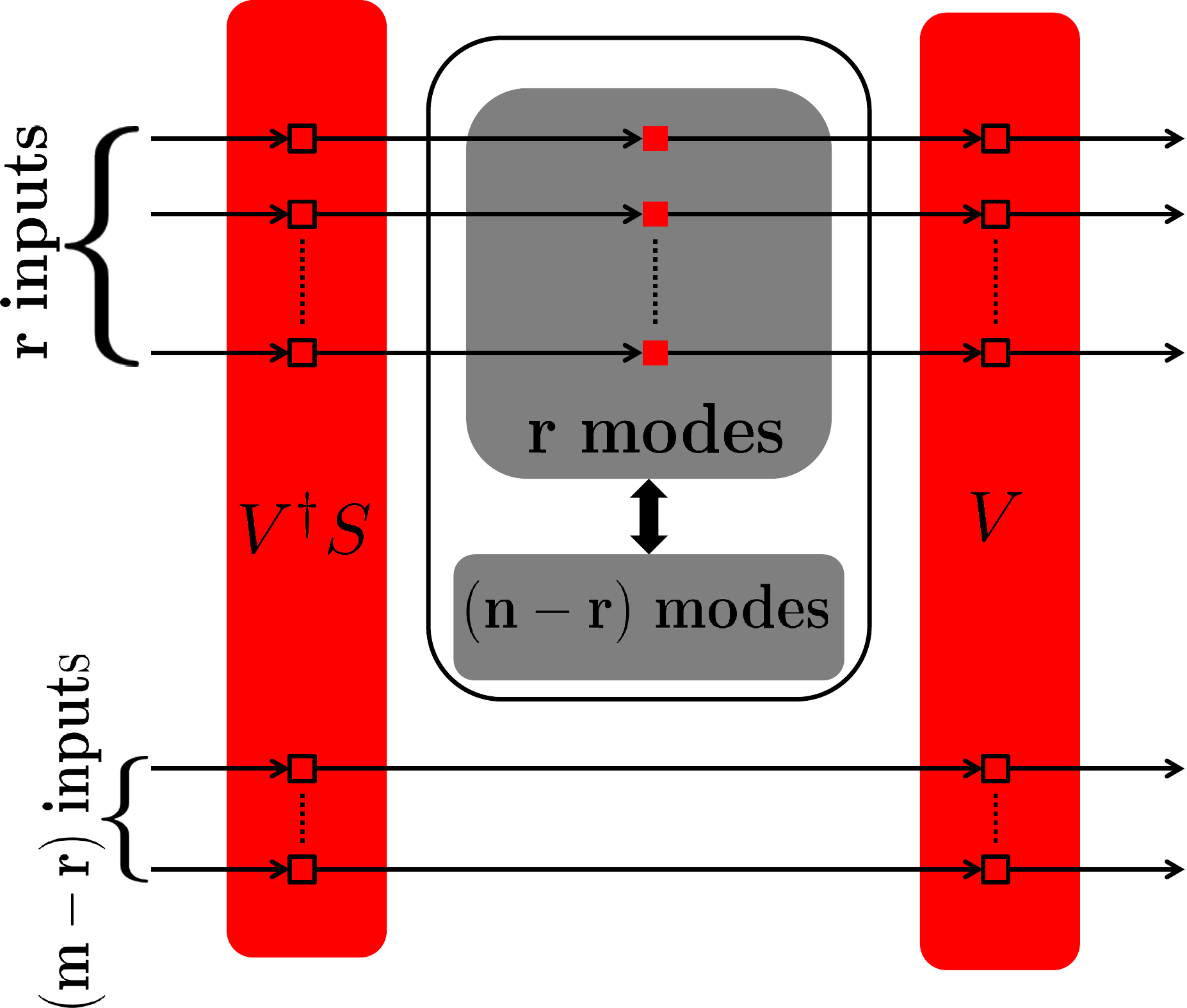}} \caption{Graphical representation of the factorization (\ref{passive tf factorization}) for a passive system. The white block represents $\hat{G}$, while the upper/lower gray blocks represent the system modes that are influenced directly/indirectly, respectively, by the inputs.} \label{Passive_Network_1}
\end{center}
\end{figure}

We first present our method of realization for transfer functions of LQSSs in the case of passive systems first, because it is the simplest case. As discussed in Subsection \ref{Linear Quantum Stochastic Systems}, a passive linear quantum stochastic system is described by the following equations:
\begin{eqnarray*}
da &=& [-\imath M -\frac{1}{2}N^{\dag}N] \,a\, dt -N^{\dag}S\, d\mathcal{U}, \\
d\mathcal{Y} &=& N a\, dt + S\, d\mathcal{U},
\end{eqnarray*}
and its transfer function is given by $ G(s)=\big[ I-N\, [sI+\imath M+\frac{1}{2}\, N^{\dag}N]^{-1}N^{\dag}\big] S$. The first step is to simplify the coupling between the system and its inputs. In order to do this, we perform the singular value decomposition (SVD) of the coupling matrix $N$, namely $N=V\hat{N}W^{\dag}$. The matrices $V$ and $W$ are unitary, and $\hat{N}$ has the following structure:
\begin{equation}\label{N_hat, passive case}
\hat{N} = \left(\begin{array}{c|c}
\bar{N}^{r \times r} & \mathbf{0} \\ \hline
\mathbf{0} & \mathbf{0} \\
\end{array}\right)=
\left(\begin{array}{ccc|c}
\sqrt{\kappa}_1 & & & \\
& \ddots & & \mathbf{0} \\
& & \sqrt{\kappa}_r & \\ \hline
& \mathbf{0} & & \mathbf{0} \\
\end{array} \right),
\end{equation}
where $r \leq \min\{n,m\}$ is the rank of $N$, and $\kappa_i >0$, $i=1,\ldots,r$. Using the SVD of $N$ in the expression for $G(s)$, and recalling that $V^{\dag}=V^{-1}$ and $W^{\dag}=W^{-1}$ (unitary matrices), we can factorize $G(s)$ as follows:
\begin{eqnarray}
G(s)= V \, \Big[I-\hat{N}\,[sI + \imath (W^{\dag}M W) + \frac{1}{2}\hat{N}^{\dag}\hat{N}]^{-1} \, \hat{N}^{\dag} \Big] (V^{\dag}S) =
V \, \hat{G}(s) \, (V^{\dag}S). \label{passive tf factorization}
\end{eqnarray}
The first and last factors in this factorization of $G(s)$, are unitary transformations of the output and the input, respectively, of the transfer function $\hat{G}(s)$ in the middle factor. As discussed in Subsection \ref{Static Linear Optical Devices and Networks}, they can be realized by multi-beam splitters. The transfer function $\hat{G}(s)$ is that of a passive linear quantum stochastic system with scattering matrix $I$, coupling matrix $\hat{N}$, and Hamiltonian matrix $\hat{M}=W^{\dag}M W$. We shall refer to this system as the reduced system associated to (\ref{Passive LQSS 2}). The structure of $\hat{N}$ is such that $r$ of the inputs of $\hat{G}(s)$ each enter into a separate port of that system and influence a corresponding (separate) mode. The remaining $m-r$ inputs ``pass through'' that system without influencing any mode. This means that $\hat{G}(s)$ is block-diagonal, with the second block being just an identity matrix:
\begin{equation}\label{Block-diagonal structure of hat_G}
\hat{G}(s)= \left(\begin{array}{c|c}
\hat{G}_{r}(s) & \mathbf{0} \\ \hline
\mathbf{0} & I_{(m-r)} \\
\end{array}\right).
\end{equation}
Also, $n-r$ of the system modes are not influenced directly by any input. This decomposition has been proposed independently in \cite{gouzha15}. The situation is depicted in Figure \ref{Passive_Network_1}.
%-------------------
%-------------------

\begin{figure}[!h]
\begin{center}
\scalebox{.3}{\includegraphics{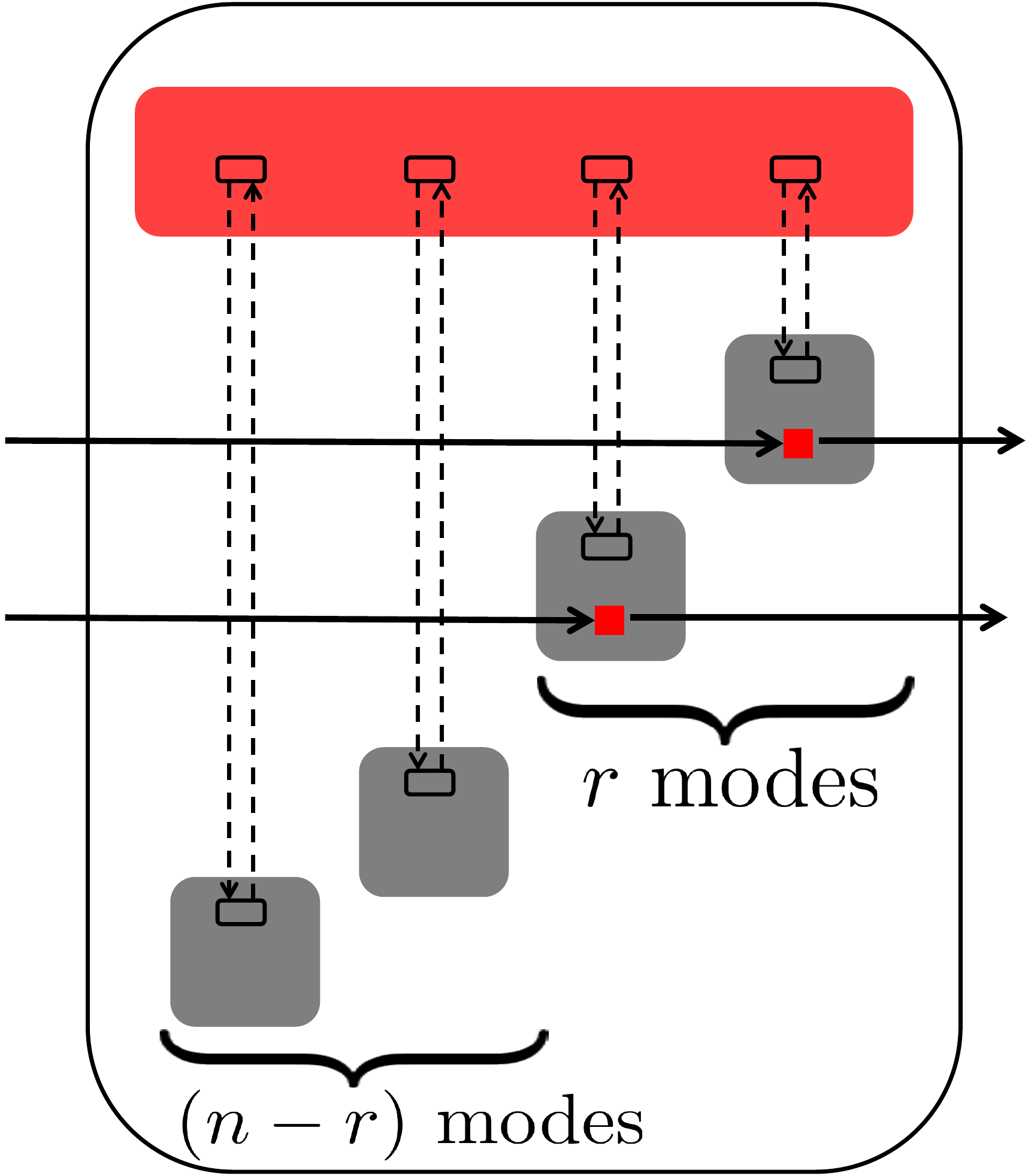}} \caption{Graphical representation of the transfer function $\hat{G}_r(s)$. The small hollow orthogonals represent passive interconnection ports.} \label{Passive_Network_2}
\end{center}
\end{figure}

\begin{figure}[!h]
\begin{center}
\scalebox{.3}{\includegraphics{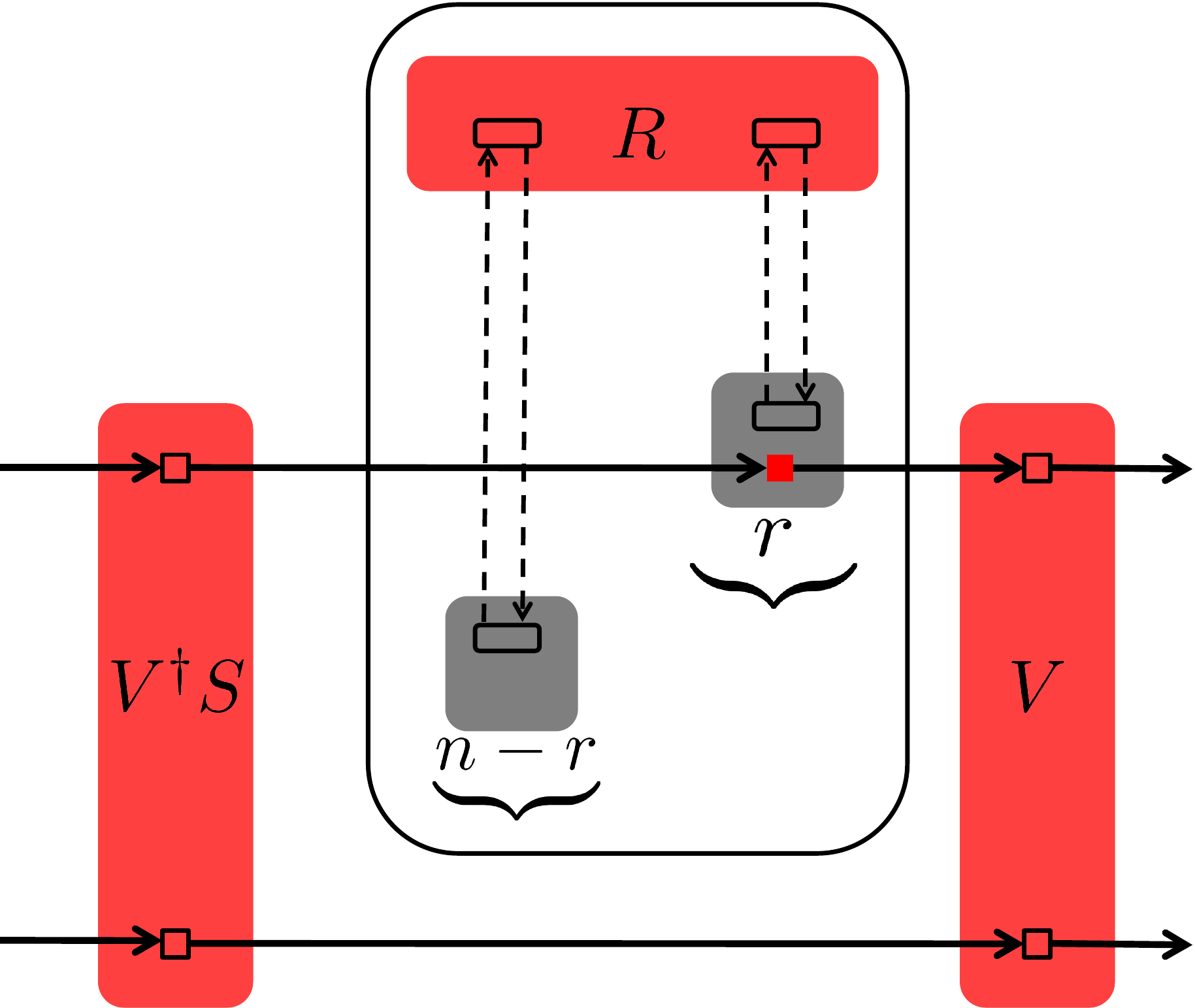}} \caption{Graphical representation of the realization method for passive LQSS transfer functions of Theorem \ref{Feedback network realization, passive case}. Each cavity represents all others of its type.} \label{Passive_Network_3}
\end{center}
\end{figure}

\begin{figure}[!h]
\begin{center}
\scalebox{.3}{\includegraphics{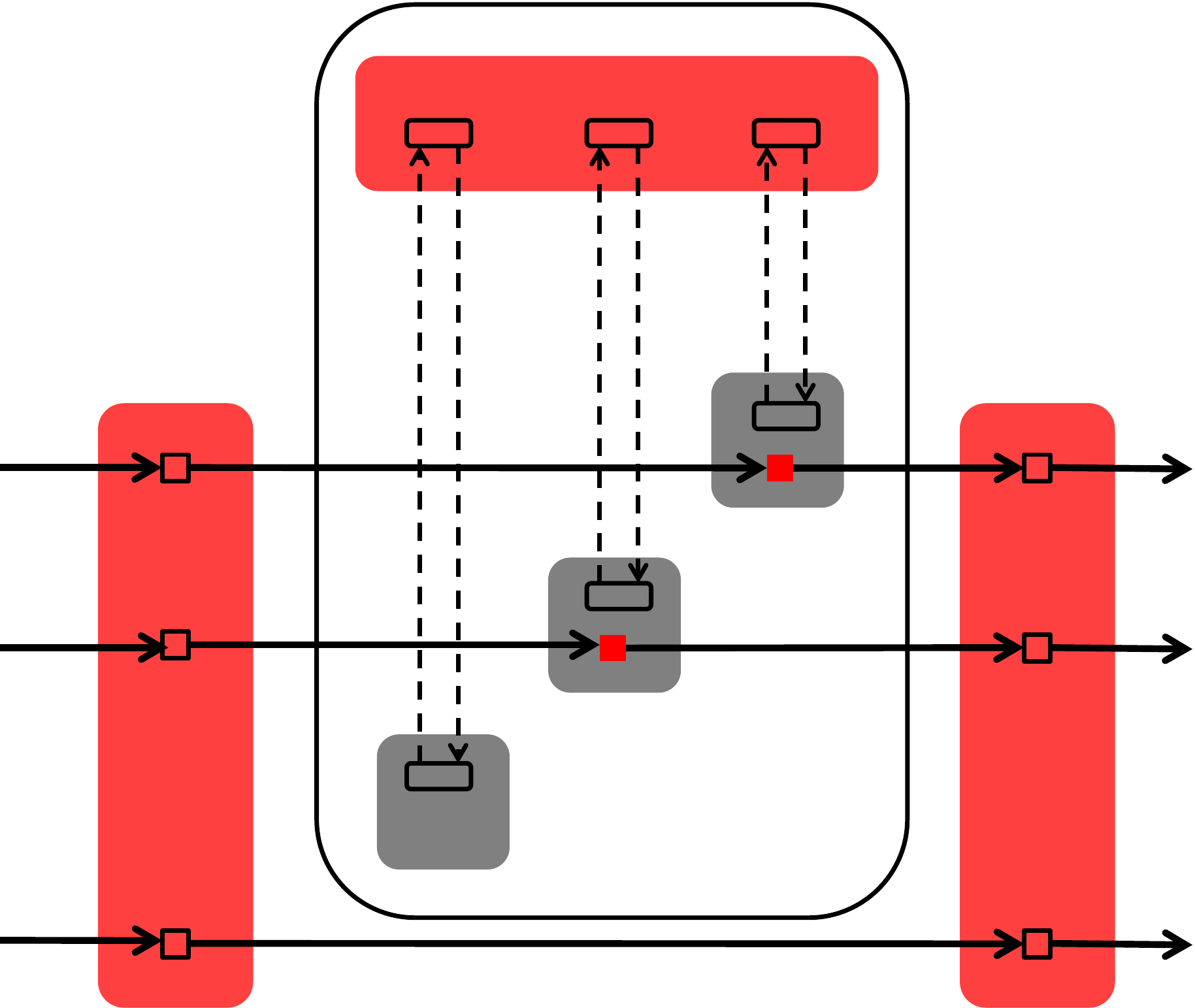}} \caption{Graphical representation of the proposed implementation of the passive transfer function in Example \ref{example passive case}.} \label{Passive_Network_4}
\end{center}
\end{figure}

Now we seek a simple representation for the reduced system
\begin{eqnarray}
da &=& [-\imath \hat{M} - \frac{1}{2}\hat{N}^{\dag}\hat{N}]\, a \, dt -\hat{N}^{\dag} d\mathcal{U}, \nonumber \\
d\mathcal{Y} &=& \hat{N} a\, dt + d\mathcal{U}, \label{reduced passive system}
\end{eqnarray}
where the same notation $a$, $\mathcal{U}$, and $\mathcal{Y}$ is used for the modes, inputs, and outputs of the reduced system, so that we do not proliferate the notation. In the reduced system each input influences at most one dynamical mode, hence, its simplest possible realization would use a collection of $n$ separate optical cavities, $r$ of which would be connected to the inputs (a 1-port cavity for each of the $r$ ``interacting'' inputs), and $n-r$ of which would not be connected to any inputs. If $\hat{M}$ were diagonal, this realization would be correct. It is apparent, however, that such a configuration would not produce the correct Hamiltonian $\hat{M}$, for general $W$ and $M$. Nevertheless, there is an easy solution to this: Each cavity should have a second port used for interconnections of the cavities through a multi-beam splitter. We show that with this feedback, we can produce any desired Hamiltonian $\hat{M}$. The model for the interconnected cavities is the following:
\begin{eqnarray}
da &=& [-\imath D - \frac{1}{2} \tilde{N}^{\dag} \tilde{N} - \frac{1}{2} \hat{N}^{\dag} \hat{N}]\, a \,dt - \tilde{N}^{\dag} d\mathcal{U}_{int}
- \hat{N}^{\dag} d\mathcal{U}, \nonumber \\
d\mathcal{Y} &=& \hat{N} a\, dt + d\mathcal{U}, \nonumber \\
d\mathcal{Y}_{int} &=& \tilde{N} a\, dt + d\mathcal{U}_{int}, \nonumber \\
d\mathcal{U}_{int} &=& R\, d\mathcal{Y}_{int}. \label{Model of cavities with interconnection}
 \end{eqnarray}
Here, $D \doteq \diag(\Delta_1,\ldots,\Delta_n)$, and $\tilde{N} \doteq \diag(\sqrt{\tilde{\kappa}_1},\ldots,\sqrt{\tilde{\kappa}_n})$, where $\Delta_i \in \mathbb{R}$, and  $\tilde{\kappa}_i>0$, are the cavity detuning and the coupling coefficient of the \emph{interconnection port} of the $i$-th cavity. The $\kappa_i>0$, $i=1,\ldots,r$ in $\hat{N}$ are the coupling coefficients of the ports of the first $r$ cavities that connect to system inputs and outputs (\emph{system ports}). The $m$-dimensional vectors $\mathcal{U}$, and $\mathcal{Y}$, contain the inputs/outputs of the system ports, and the $n$-dimensional vectors $\mathcal{U}_{int}$, and $\mathcal{Y}_{int}$, the inputs/outputs of the interconnection ports.
$R$ is the $n \times n$ unitary transformation that is implemented by a multi-beam splitter that introduces interconnections between the outputs and the inputs of the interconnection ports of the cavities, see Figure \ref{Passive_Network_2}.
%-------------------
%-------------------
Combining the last two equations in (\ref{Model of cavities with interconnection}), we obtain the relation $d\mathcal{U}_{int} = (I-R)^{-1} R\, \tilde{N} a\, dt$. At this point we introduce a variant of the Cayley transform for unitary matrices without unit eigenvalues \cite{golvan96}, namely
\begin{equation}\label{Cayley transform}
X=(I-R)^{-1}(I+R).
\end{equation}
The unitarity of $R$ implies that $X$ is skew-Hermitian. We can also solve uniquely for $R$ in terms of $X$ with the following result:
\begin{equation}\label{inverse Cayley transform}
R=(X-I)(X+I)^{-1},
\end{equation}
where $R$ is defined for all skew-Hermitian matrices $X$, and can be seen to be unitary due to the skew-Hermitian nature of $X$. So, this map from the $n$-dimensional unitary matrices without unit eigenvalues to the $n$-dimensional skew-Hermitian matrices, is 1-1 and onto. Tt is easy to see that $(I-R)^{-1}R= - \frac{1}{2} I + \frac{1}{2}X$. Using the relation between $d\mathcal{U}_{int}$ and $a$, and the definition of $X$, the equations for the network take the following form:
\begin{eqnarray}
da &=& [-\imath D - \frac{1}{2} \tilde{N}^{\dag} X \tilde{N} - \frac{1}{2} \hat{N}^{\dag} \hat{N}]\, a \,dt - \hat{N}^{\dag} d\mathcal{U} ,  \nonumber \\
d\mathcal{Y} &=& \hat{N} a\, dt + d\mathcal{U}.  \label{interconnected cavities model}
\end{eqnarray}
These equations describe a passive linear quantum stochastic system with Hamiltonian matrix $\hat{M}$ given by the expression
\begin{equation}\label{Network Hamiltonian}
\hat{M}=D - \frac{\imath}{2} \tilde{N}^{\dag} X \tilde{N}.
\end{equation}
Given any values for the cavity parameters $\Delta_i$ and $\tilde{\kappa}_i >0$, and any desired Hamiltonian matrix $\hat{M}=W^{\dag}M W$, we may determine the unique $X$ (and hence the unique $R$) that achieves this $\hat{M}$ by the expression
\begin{equation}\label{solution for X}
X=2\imath \tilde{N}^{-\dag} (\hat{M}-D) \tilde{N}^{-1}.
\end{equation}
We summarize the proposed methodology in the following theorem:
\begin{theorem}\label{Feedback network realization, passive case}
Given a passive linear quantum stochastic system with Hamiltonian matrix $M^{n \times n}$, coupling operator $N^{m \times n}$, and scattering matrix $S^{m \times m}$, let
\[ G(s)=S-N\, [sI+\imath M+\frac{1}{2}\, N^{\dag}N]^{-1}N^{\dag}S \]
be its transfer function. Let $N=V\hat{N}W^{\dag}$ be the singular value decomposition of the coupling matrix $N$. Then, $G(s)$ can be factorized as $G(s)= V\,\hat{G}(s)\, (V^{\dag}S)$, where $\hat{G}(s)$ has the form
\[ \hat{G}(s)= I-\hat{N}\,[sI + \imath (W^{\dag}M W) + \frac{1}{2}\hat{N}^{\dag}\hat{N}]^{-1} \, \hat{N}^{\dag}. \]
Moreover, $\hat{G}(s)$ may be realized by the following feedback network of $(n-r)$ 1-port and $r$ 2-port passive cavities:
\begin{eqnarray*}
da &=& [-\imath D - \frac{1}{2} \tilde{N}^{\dag} \tilde{N} - \frac{1}{2} \hat{N}^{\dag} \hat{N}]\, a \,dt - \tilde{N}^{\dag} d\mathcal{U}_{int}
- \hat{N}^{\dag} d\mathcal{U}, \\
d\mathcal{Y} &=& \hat{N} a\, dt + d\mathcal{U}, \\
d\mathcal{Y}_{int} &=& \tilde{N} a\, dt + d\mathcal{U}_{int}, \\
d\mathcal{U}_{int} &=& R\, d\mathcal{Y}_{int}.
 \end{eqnarray*}
Here, $D = \diag(\Delta_1,\ldots,\Delta_n)$, and $\tilde{N} = \diag(\sqrt{\tilde{\kappa}_1},\ldots,\sqrt{\tilde{\kappa}_n})$, where $\Delta_i \in \mathbb{R}$, and  $\tilde{\kappa}_i>0$, are the cavity detuning and the coupling coefficient of the interconnection port, respectively, of the $i$-th cavity. The $m$-dimensional vectors $\mathcal{U}$, and $\mathcal{Y}$, contain the inputs/outputs of the system ports, and the $n$-dimensional vectors $\mathcal{U}_{int}$, and $\mathcal{Y}_{int}$, the inputs/outputs of the interconnection ports. Finally, the unitary interconnection matrix (feedback gain) $R$ is determined through the relations
\begin{eqnarray*}
X &=& 2\imath \tilde{N}^{-\dag} (W^{\dag}M W-D) \tilde{N}^{-1}, \\
R &=& (X-I)(X+I)^{-1}.
\end{eqnarray*}
\end{theorem}
Figure \ref{Passive_Network_3} provides a graphical representation of the realization method of Theorem \ref{Feedback network realization, passive case}.
%-------------------
%-------------------
We end this section with an illustrative example.
\begin{example}\label{example passive case}
Consider the 3-mode, 3-input passive linear quantum stochastic system with the following parameters:
\begin{equation*}
M = \left(\begin{array}{rrr}
5&1&-2 \\
1&3&0 \\
-2&0&4 \\
\end{array}\right), \
N = \left(\begin{array}{rrr}
1&2&1 \\
0&-1&3 \\
2&3&5 \\
\end{array}\right),\ \mathrm{and,}\ S=I_3 .
\end{equation*}
The SVD of $N$ is given by $N=V\hat{N}W^{\dag}$, with
\begin{eqnarray*}
V &=& \left(\begin{array}{rrr}
-0.2987 & 0.4941 & -0.8165\\
-0.3065 & -0.8599 & -0.4082\\
-0.9038 & 0.1283 & 0.4082\\
\end{array}\right),\\
W &=& \left(\begin{array}{rrr}
-0.3093 & 0.2717 & -0.9113\\
-0.4409 & 0.8081 & 0.3906\\
-0.8426 & -0.5226 & 0.1302\\
\end{array}\right),\ \mathrm{and,} \\
\hat{N} &=& \diag( 6.8092, 2.7632, 0).
\end{eqnarray*}
The Hamiltonian of the reduced system is given by
\begin{equation*}
\hat{M}=W^{\dag}M W = \left(\begin{array}{rrr}
3.1315 & 0.0370 & -0.7200\\
0.0370 & 4.4278 & -2.2169\\
-0.7200 & -2.2169 & 4.4407\\
\end{array}\right).
\end{equation*}
Letting $D=0_{3 \times 3}$ and $\tilde{N}=I_3$, equation (\ref{solution for X}) produces the following $X$:
\begin{equation*}
X= \imath \left(\begin{array}{rrr}
6.2631 &  0.0740 & -1.4400 \\
0.0740 &  8.8556 & -4.4337 \\
-1.4400 & -4.4337 &  8.8814 \\
\end{array}\right),
\end{equation*}
from which we calculate the feedback gain matrix $R$ using equation (\ref{inverse Cayley transform}),
\begin{equation*}
R= \left(\begin{array}{rrr}
0.9429 & -0.0145 & -0.0237 \\
-0.0145 & 0.9438 & -0.0467 \\
-0.0237 & -0.0467 & 0.9389 \\
\end{array} \right)
+ \imath \left(\begin{array}{rrr}
0.3245 & 0.0276 & 0.0637 \\
0.0276 & 0.2918 & 0.1449 \\
0.0637 & 0.1449 & 0.3010 \\
\end{array} \right)
\end{equation*}
\end{example}
Figure \ref{Passive_Network_4} provides a graphical representation of the proposed implementation of the transfer function for this example.
%-------------------
%-------------------

\section{Realization of General Linear Quantum Stochastic Systems}
\label{Realization of General Linear Quantum Stochastic Systems}

In this section, we present our synthesis method for the case of a general linear quantum stochastic system. As described in Section \ref{Background Material}, the model for such a system is the following:
\begin{eqnarray*}
d\check{a} &=& [-\imath JM -\frac{1}{2}N^{\flat}N]\, \check{a} dt - N^{\flat} S d\check{\mathcal{U}}, \\
d\check{\mathcal{Y}} &=& N \check{a} dt + S d\check{\mathcal{U}},
\end{eqnarray*}
with transfer function from $d\check{\mathcal{U}}$ to $d\check{\mathcal{Y}}$ given by $G(s)= \big[I-N [sI+\imath JM +\frac{1}{2}N^{\flat}N]^{-1} N^{\flat} \big] S$. To proceed as in Section \ref{Realization of Passive Linear Quantum Stochastic Systems}, we derive two results. First, we derive a canonical form for doubled-up matrices which generalizes the usual SVD that we employed in Section \ref{Realization of Passive Linear Quantum Stochastic Systems}. This result is along the following lines: Given a complex doubled-up matrix $N^{2m \times 2n}$, there exist Bogoliubov matrices $V^{2m \times 2m}$ and $W^{2n \times 2n}$, and a doubled-up matrix $\hat{N}^{2m \times 2n}$ in a standard reduced form (to be specified in the following), such that $N = V \, \hat{N} \, W^{\flat}$. Using this factorization of $N$ in the expression for $G(s)$, along with the fact that $V^{\flat}=V^{-1}$, and $W^{\flat}=W^{-1}$, $G(s)$ can be factorized as follows:
\begin{eqnarray}
G(s) &=& V \, \Big[I-\hat{N}\, [sI+ \imath J (W^{\dag}M W) +\frac{1}{2}\hat{N}^{\flat}\hat{N}]^{-1} \hat{N}^{\flat} \Big]\, (V^{\flat}S) = V \, \hat{G}(s)\, (V^{\flat} S). \label{general tf factorization}
\end{eqnarray}
The first and last factors in this factorization of $G(s)$, are Bogoliubov transformations of the output and the input, respectively, of the transfer function $\hat{G}(s)$ in the middle factor. As discussed in Subsection \ref{Static Linear Optical Devices and Networks}, they can be realized by multi-squeezers. The transfer function $\hat{G}(s)$, is that of a linear quantum stochastic system with generalized scattering matrix $I$, coupling matrix $\hat{N}$, and Hamiltonian matrix $\hat{M}=W^{\dag}M W$. We shall refer to it as the reduced system associated to (\ref{General LQSS2}). In Section \ref{Realization of Passive Linear Quantum Stochastic Systems}, we saw that the structure of the $\hat{N}$ matrix suggested the use of passive cavities as the simplest dynamical elements to realize the associated reduced system. At this point we have not yet specified the structure of the coupling matrix $\hat{N}$ in the case of general systems, hence, we cannot propose yet the types of devices needed to implement it. Nevertheless, it is obvious that we shall need a second result along the following lines: Given any desired Hamiltonian matrix $\hat{M}$ for the reduced system, we can obtain it from the Hamiltonian $M_{conc}$ of the collection (concatenation) of devices used to realize the reduced system, with appropriate feedback through a multi-squeezer.

\begin{figure}[!h]
\begin{center}
\scalebox{.4}{\includegraphics{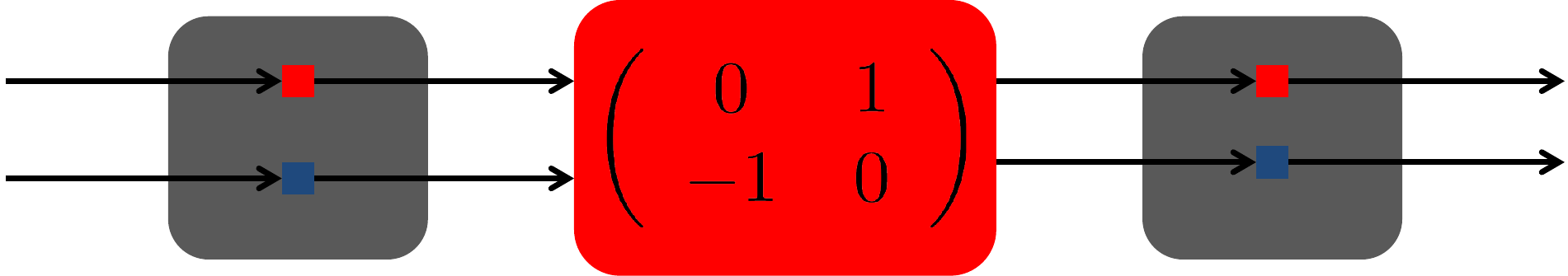}} \caption{A simple cascade system realizing $\hat{N}$ in the case of a complex eigenvalue.} \label{Complex_eigenvalue_block_decomposition}
\end{center}
\end{figure}

We now state the aforementioned results precisely, and prove them. We begin with an SVD type of result (canonical form) for doubled-up matrices:
\begin{theorem} \label{Bogoliubov SVD}
Let $N^{2m \times 2n}=\bigl(\begin{smallmatrix} N_1 & N_2 \\ N_2^{\#} & N_1^{\#} \end{smallmatrix}\bigr)$ be a complex doubled-up matrix, and let $\mathcal{N} \doteq N^{\flat}N$. We assume that all the eigenvalues of $\mathcal{N}$ are semisimple. Also, we assume that $\Ker \mathcal{N}=\Ker N$, i.e all the eigenvectors of $\mathcal{N}$ with zero eigenvalue belong to the kernel of $N$. Let $\lambda_i^+ >0, i=1,\ldots,r_+$, $\lambda_i^- <0, i=1,\ldots,r_-$, and $\lambda_i^c$, with $\Im\lambda_i^c >0$, $i=1\ldots,r_c$, be the eigenvalues of $\mathcal{N}$ that are, respectively, positive, negative, and non-real with positive imaginary part. Then, there exist Bogoliubov matrices $V^{2m \times 2m}$, $W^{2n \times 2n}$, and a complex matrix $\hat{N}^{2m \times 2n}=\bigl(\begin{smallmatrix} \hat{N}_1 & \hat{N}_2 \\ \hat{N}_2^{\#} & \hat{N}_1^{\#} \end{smallmatrix}\bigr)$, such that $N = V \, \hat{N} \, W^{\flat}$, where $\hat{N}_{1}= \bigl(\begin{smallmatrix} \bar{N}_{1} & \mathbf{0} \\ \mathbf{0} & \mathbf{0} \end{smallmatrix}\bigr)$, $\hat{N}_{2}= \bigl(\begin{smallmatrix} \bar{N}_{2} & \mathbf{0} \\ \mathbf{0} & \mathbf{0} \end{smallmatrix}\bigr)$, and
\begin{eqnarray*}
\bar{N}_1^{r \times r} &=& \diag(\sqrt{\lambda_1^+},\ldots,\sqrt{\lambda_{r_+}^+},\underbrace{0,\ldots,0}_{r_-},\alpha_1 I_2,\ldots,\alpha_{r_c} I_2), \\
\bar{N}_2^{r \times r} &=& \diag(\underbrace{0,\ldots,0}_{r_+},\sqrt{|\lambda_1^-|},\ldots,\sqrt{|\lambda_{r_-}^-|},-\beta_1\sigma_2,\ldots,-\beta_{r_c}\sigma_2),
\end{eqnarray*}
where $r=r_+ + r_- + 2r_c$, and $\sigma_2 = \bigl(\begin{smallmatrix} 0 & -\imath \\ \imath & 0 \end{smallmatrix} \bigr)$ is one of the Pauli matrices. The parameters $\alpha_i$ and $\beta_i$ are determined in terms of $\lambda_i^c$, as follows:
\[ \alpha_i = \sqrt{\frac{|\lambda_i^c|+\Re \lambda_i^c}{2}}, \ \ \beta_i = \frac{\Im \lambda_i^c}{\sqrt{2\big(|\lambda_i^c|+\Re \lambda_i^c\big)}}. \]
\end{theorem}
The proof of the theorem is presented in the appendix, along with some remarks extending its applicability to a larger class of matrices than announced in its statement. According to the theorem, the simplest possible forms of $\hat{N}$ are the following:
\begin{itemize}
\item For a positive eigenvalue $\lambda$ of $\mathcal{N}$, $\hat{N}=\bigl(\begin{smallmatrix} \sqrt{\lambda} & 0 \\ 0 & \sqrt{\lambda} \end{smallmatrix}\bigr)$. The simplest implementation would be with a cavity with a passive port of coefficient $\lambda$.
\item For a negative eigenvalue $\lambda$ of $\mathcal{N}$, $\hat{N}=\bigl(\begin{smallmatrix} 0 & \sqrt{|\lambda|} \\ \sqrt{|\lambda|} & 0 \end{smallmatrix}\bigr)$. The simplest implementation would be with a cavity with an active port of coefficient $|\lambda|$.
\item For a non-real eigenvalue $\lambda$,
\[ \hat{N}= \left(\begin{array}{rr} \alpha I_2 & -\beta \sigma_2 \\ \beta \sigma_2 & \alpha I_2 \\ \end{array}\right)=
\left(\begin{array}{rr|rr}
\alpha & 0 & 0 & \imath \beta \\
0 & \alpha & -\imath\beta & 0 \\ \hline
0 & -\imath\beta & \alpha & 0 \\
\imath\beta & 0 & 0 & \alpha  \\ \end{array}\right), \]
where $\alpha$ and $\beta$ are given by the corresponding expressions at the end of Theorem \ref{Bogoliubov SVD}. It is straightforward to verify that this $\hat{N}$ can be implemented by the cascade connection of two identical 2-port cavities and a beam-splitter, as in Figure \ref{Complex_eigenvalue_block_decomposition}. The cavity has two ports, one passive with coupling coefficient $\alpha^2$, and one purely active with coupling coefficient $\beta^2$. Its coupling matrix $N_c$ is given by
\[ N_c= \left(\begin{array}{rr}
0 & \imath\beta \\
\alpha & 0 \\
-\imath\beta & 0 \\
0 & \alpha \\
\end{array}\right). \]
The beam splitter implements the unitary transformation $\bigl(\begin{smallmatrix} 0 & 1 \\ -1 & 0 \end{smallmatrix}\bigr)$. If the Hamiltonian matrix of each cavity is given by $M_c= \bigl(\begin{smallmatrix} \Delta & 0 \\ 0 & \Delta \end{smallmatrix}\bigr)$, the total Hamiltonian matrix of the two-cavity system is given by
\[M= \left(\begin{array}{cccc}
\Delta & 0 & 0 & -\Im\lambda/2 \\
0 & \Delta & -\Im\lambda/2 & 0 \\
0 & -\Im\lambda/2 & \Delta & 0 \\
-\Im\lambda/2 & 0 & 0 & \Delta \\
\end{array}\right).\]
\end{itemize}
%
%-------------------
%-------------------

We turn our attention to the second result necessary in our synthesis method. We show that given a collection of quantum optical dynamical devices that implement the desired reduced coupling matrix $\hat{N}$, their collective Hamiltonian matrix can be altered to produce any desired Hamiltonian matrix using feedback through a multi-squeezer. In fact, we prove a more general statement:
\begin{theorem}\label{Feedback modification of a LQSS Hamiltonian}
Given a linear quantum stochastic system described by the model
\begin{eqnarray*}
d\check{a} &=& [-\imath JM -\frac{1}{2}N^{\flat}N]\, \check{a} dt - N^{\flat} S d\check{\mathcal{U}}, \\
d\check{\mathcal{Y}} &=& N \check{a} dt + S d\check{\mathcal{U}},
\end{eqnarray*}
consider the modified system
\begin{eqnarray}
d\check{a} &=& [-\imath J M -\frac{1}{2}\tilde{N}^{\flat}\tilde{N} - \frac{1}{2} N^{\flat} N]\, \check{a} dt
- \tilde{N}^{\flat} d\check{\mathcal{U}}_{int} - N^{\flat} S d\check{\mathcal{U}}, \nonumber \\
d\check{\mathcal{Y}} &=& N \check{a} dt + S d\check{\mathcal{U}}, \nonumber \\
d\check{\mathcal{Y}}_{int} &=& \tilde{N} \check{a} dt + d\check{\mathcal{U}}_{int} , \nonumber \\
d\check{\mathcal{U}}_{int} &=& R\, d\check{\mathcal{Y}}_{int}, \label{Model of general system with interconnection}
\end{eqnarray}
with $\tilde{N}=\diag(\sqrt{\tilde{\kappa}_1},\ldots,\sqrt{\tilde{\kappa}_n}, \sqrt{\tilde{\kappa}_1},\ldots,\sqrt{\tilde{\kappa}_n})$ ($\tilde{\kappa}_i >0$), and $R$ a Bogoliubov matrix. The new system is constructed from the original one by adding $n$ passive interconnection ports (one for every mode), and feeding back through a multi-squeezer. The $m$-dimensional vectors $\mathcal{U}$, and $\mathcal{Y}$, contain the inputs/outputs of the original system ports, and the $n$-dimensional vectors $\mathcal{U}_{int}$, and $\mathcal{Y}_{int}$, the inputs/outputs of the interconnection ports. Then, there is always a unique $R$ such that the modified system has any desired Hamiltonian matrix $\bar{M}$.
\end{theorem}
\textbf{Proof:} Combining the last two equations of (\ref{Model of general system with interconnection}), we obtain the expression $ d\check{\mathcal{U}}_{int} = R(I-R)^{-1} \tilde{N} \check{a} dt$. As in Section \ref{Realization of Passive Linear Quantum Stochastic Systems}, we introduce the Cayley transform $X=(I+R)(I-R)^{-1}$, defined for Bogoliubov matrices $R$ with no unit eigenvalues. Its unique inverse is defined by $R=(X-I)(X+I)^{-1}$. It is straightforward to verify that, $X$ is doubled-up and $\flat$-skew-Hermitian ($X^{\flat}=-X$) if and only if $R$ is Bogoliubov. Using the identity $R(I-R)^{-1}= - \frac{1}{2} I + \frac{1}{2}X$, we reduce the model of the modified system as follows:
\begin{eqnarray*}
d\check{a} &=& [-\imath J M -\frac{1}{2}\tilde{N}^{\flat} X \tilde{N} - \frac{1}{2} N^{\flat} N]\, \check{a} dt - N^{\flat} S d\check{\mathcal{U}},  \\
d\check{\mathcal{Y}} &=& N \check{a} dt + S d\check{\mathcal{U}}.
 \end{eqnarray*}
The new Hamiltonian $\bar{M}$ is given by the expression $J\bar{M}= JM-\frac{\imath}{2} (\tilde{N}^{\flat} X \tilde{N})$, which can be solved uniquely for the matrix $X$ that produces the desired Hamiltonian, given the parameters $\bar{M}$, and $\tilde{N}$, namely $X=2\imath (\tilde{N}^{\flat})^{-1} (J\bar{M}-JM)\, \tilde{N}^{-1}$. So, the corresponding $R$ is determined uniquely by the inverse Cayley transform.$\blacksquare$

\begin{figure}[!h]
\begin{center}
\scalebox{.4}{\includegraphics{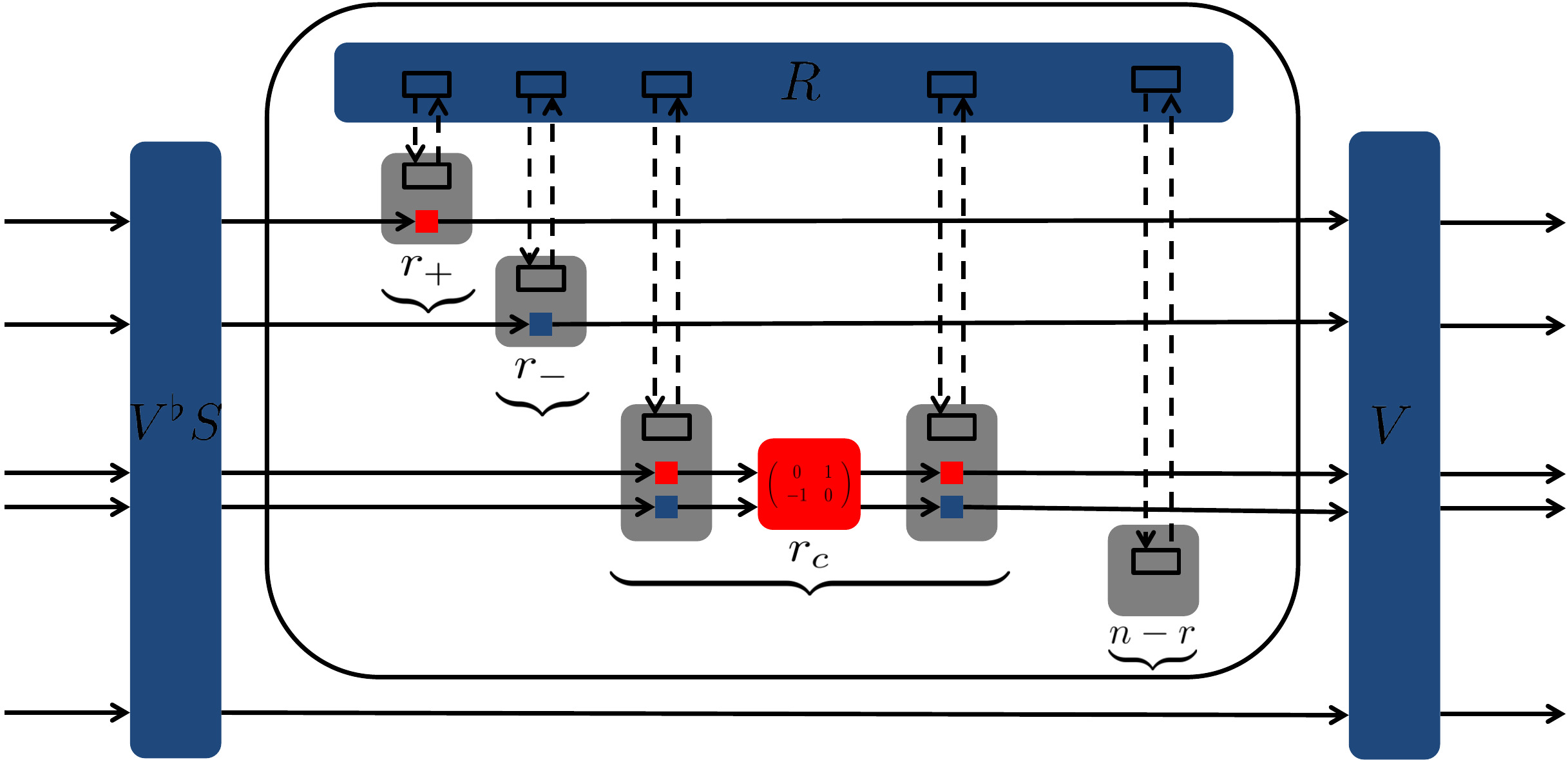}} \caption{A graphical representation of the realization of the transfer function of a general LQSS. Each cavity is representative of all cavities of its type needed to implement the transfer function.} \label{Active_Network_1}
\end{center}
\end{figure}

Let $M_{conc}$ be the Hamiltonian matrix of the collection (concatenation) of quantum optical dynamical devices that implement the desired reduced coupling matrix $\hat{N}$. The application of Theorem \ref{Feedback modification of a LQSS Hamiltonian} with $M$ being $M_{conc}$, and $\bar{M}=\hat{M}=W^{\dag}M W$, completes the synthesis. Figure \ref{Active_Network_1} is a graphical representation of the realization of the transfer function of a general LQSS. Each cavity is representative of all cavities of its type needed to implement the transfer function. Finally, we demonstrate our method with an example.
%-------------------
%-------------------
%
\begin{example}\label{example general case}
Consider the 2-mode, 2-input linear quantum stochastic system with the following parameters:
\[ M = \left(\begin{array}{rrrr}
2  &   1  &   0   &  -1\\
1  &   2  &  -1   &  0\\
0  &  -1  &   2   &  1\\
-1 &   0  &   1   &  2\\
\end{array}\right), \
N = \left(\begin{array}{rrrr}
0&1&2&0\\
-1&2&1&-1\\
2&0&0&1\\
1&-1&-1&2\\
\end{array}\right), \]
and $S=I_4$. The eigenvalue decomposition of $\mathcal{N}=N^{\flat}N$ is computed to be $\mathcal{N}=U D U^{-1}$, where $D=\diag(-2.8284,2.8284,-2.8284,2.8284)$ and
\[ U=\left(\begin{array}{rrrr}
-0.9074  &  0.3474  &  0.2038  &  0.1756\\
-0.1329  &  0.2965  &  0.4090  & -0.8908\\
0    &     0 &  -0.8629 &   0.4064\\
0.3987 &  -0.8896 &  -0.2159 &  -0.1027\\
\end{array}\right).\]
To the positive eigenvalue $\lambda^+=2.8284$, there correspond the eigenvectors $u_2$ and $u_4$ given by the second and fourth columns of $U$. We have that $\langle u_4,u_4\rangle_J >0$, and after normalization $u_4$ becomes $z^{+}=(0.2180,-1.1061,0.5046,-0.1275)^{\top}$. To the negative eigenvalue $\lambda^-=-2.8284$, there correspond the eigenvectors $u_1$ and $u_3$ given by the first and third columns of $U$. We have that $\langle u_1,u_1\rangle_J >0$, and after normalization $u_1$ becomes $z^{-}=(-1.0987,-0.1609,0,0.4827)^{\top}$. According to the proof of Theorem \ref{Bogoliubov SVD},
\[ W=\big[ [z^+ z^-]\, \Sigma\,[z^+ z^-]^{\#}\big]=
\left(\begin{array}{rrrr}
0.2180 &  -1.0987  &  0.5046   &      0\\
-1.1061 &   -0.1609  &   -0.1275 &   0.4827\\
0.5046   &    0  &  0.2180 &  -1.0987\\
-0.1275  &  0.4827 &  -1.1061 &  -0.1609\\
\end{array}\right). \]
Since there are no zero eigenvalues,
\[\hat{N}=\bar{N}=
\left(\begin{array}{rrrr}
1.6818   &      0    &     0    &     0\\
0    &     0    &     0  &   1.6818\\
0    &     0  &  1.6818    &     0\\
0  &  1.6818   &      0    &     0\\
\end{array}\right), \]
and we can compute $V$ simply by
\[ V= N\, W\, \hat{N}^{-1} =
\left(\begin{array}{rrrr}
-0.0576  & -1.0196  &  0.1834  & -0.0957\\
-1.0691  &  0.0164  &  0.3357  &  0.1749\\
0.1834  & -0.0957  &  -0.0576  & -1.0196\\
0.3357  &  0.1749 &  -1.0691  &   0.0164\\
\end{array}\right). \]
The Hamiltonian of the reduced system should be equal to
\[ \hat{M}=W^{\dag}M W =
\left(\begin{array}{rrrr}
3.6444  &  1.0135  &  0.4429  & -3.3952\\
1.0135  &  4.3462  & -3.3952  & -1.7249\\
0.4429  & -3.3952  &  3.6444  &  1.0135\\
-3.3952 &  -1.7249  & 1.0135  &  4.3462\\
\end{array}\right).  \]
The reduced system can be implemented by the use of two cavities, one with a passive port (corresponding to $\lambda^+$), and one with an active port (corresponding to $\lambda^-$). Choosing the detuning of both cavities to be zero, makes the total Hamiltonian of their concatenation $M_{conc}=0_{4 \times 4}$. Also, we choose $\tilde{N}=I_4$.
Then, we compute
\begin{eqnarray*}
X &=& 2\imath (\tilde{N}^{\flat})^{-1} J\, (\hat{M}-M)\, \tilde{N}^{-1} \\
&=&
\imath \left(\begin{array}{rrrr}
7.2889  &  2.0271  &  0.8858  &  -6.7904\\
2.0271  &  8.6924  & -6.7904  &  -3.4497\\
-0.8858 &   6.7904 &  -7.2889 &  -2.0271\\
6.7904  &  3.4497  &  -2.0271 &  -8.6924\\
\end{array}\right),
\end{eqnarray*}
from which the feedback gain $R$ is computed to be
\begin{eqnarray*}
R = (X-I)(X+I)^{-1} &=& \,\,\, \left(\begin{array}{rrrr}
-0.3731  &  0.9082  & 0        & 0.0450 \\
0.9082   &  0.3125  & -0.0450  & 0      \\
0        &  0.0450  & -0.3731  & 0.9082 \\
-0.0450  &  0       & 0.9082   & 0.3125 \\
\end{array} \right) \\
&+& \imath \left(\begin{array}{rrrr}
7.8624 &  - 5.2659 & 7.4743 & - 5.8003\\
- 5.2659  &   4.4401 & - 5.8003 &  3.7042\\
- 7.4743  &   5.8003 & - 7.8624 &  5.2659\\
5.8003 &  - 3.7042 &  5.2659 & - 4.4401\\
\end{array} \right)
\end{eqnarray*}
Figure \ref{Active_Network_2} provides a graphical representation of the proposed implementation of the transfer function for this example.
%-------------------
\begin{figure}[!h]
\begin{center}
\scalebox{.3}{\includegraphics{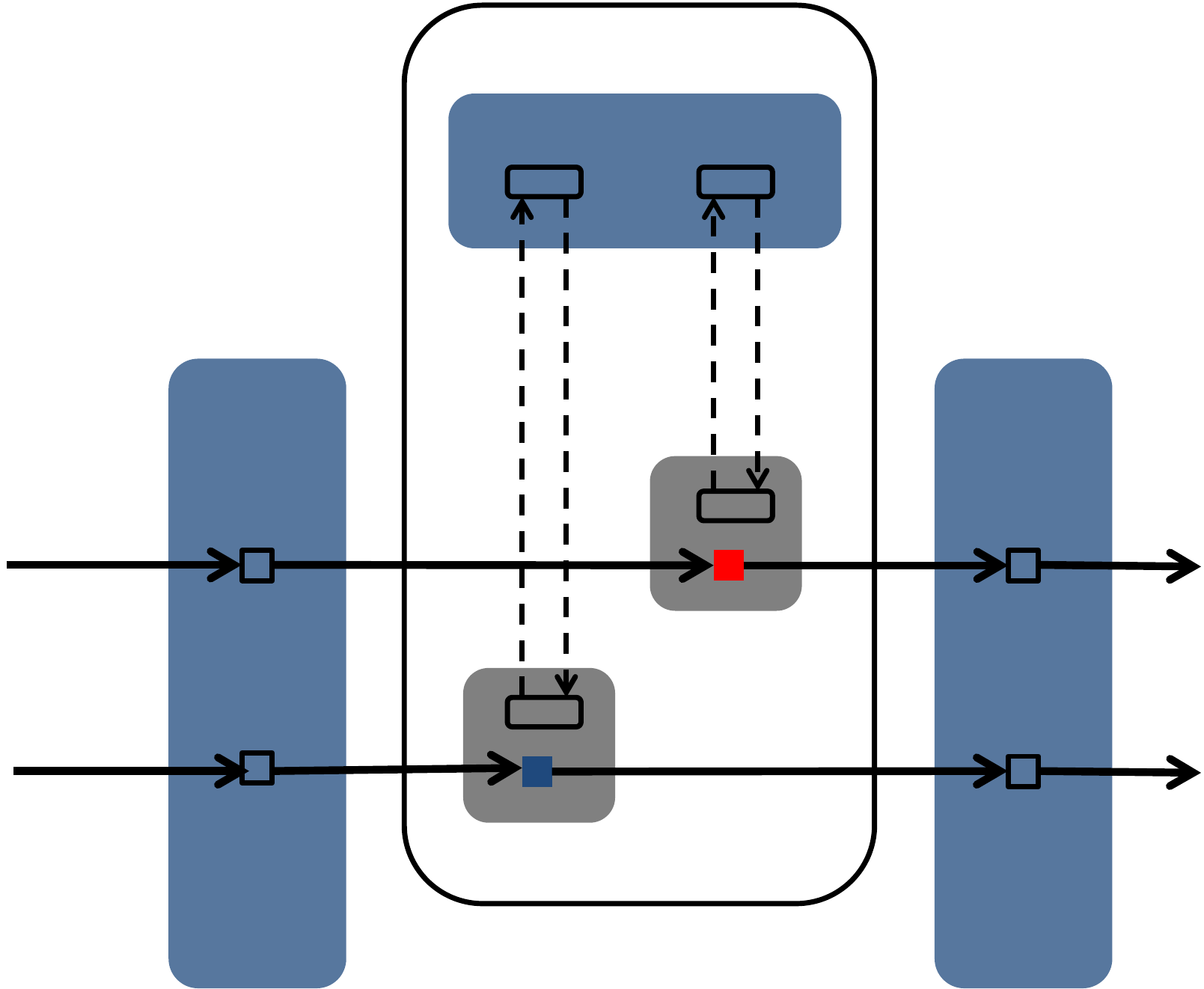}} \caption{Graphical representation of the proposed implementation of the transfer function of Example \ref{example general case}.} \label{Active_Network_2}
\end{center}
\end{figure}
\end{example}
We end this section and the paper with some remarks.
\begin{enumerate}
  \item In the case of a negative eigenvalue $\lambda$ of $\mathcal{N}$, the form of $\hat{N}$ given in Theorem \ref{Bogoliubov SVD} is  $\hat{N}=\bigl(\begin{smallmatrix} 0 & \sqrt{|\lambda|} \\ \sqrt{|\lambda|} & 0 \end{smallmatrix}\bigr)$. The simplest implementation of this $\hat{N}$ is by a cavity with a purely active port. However, there is a more general form for $\hat{N}$, that allows for the presence of damping in the port. It is given by the expression
\[ \hat{N}=\left(\begin{array}{cc} \sqrt{|\lambda|}\sinh x & \sqrt{|\lambda|}\cosh x \\ \sqrt{|\lambda|} \cosh x & \sqrt{|\lambda|}\sinh x \\ \end{array}\right),
x \in \mathbb{R}. \]
  \item Theorem \ref{Bogoliubov SVD} excludes the case of non-semisimple eigenvalues of $\mathcal{N}=N^{\flat}N$ (Jordan blocks of dimension greater than one). We point out that there is no fundamental issue in this case. In Remark 1 following the proof of Theorem \ref{Bogoliubov SVD} in the Appendix, we extend the theorem in the case of a real eigenvalue with Jordan block of dimension $2$. In principle, we could also extend the theorem in the case of real and non-real eigenvalues whose Jordan blocks are of dimension greater than two. The issue is one of complexity: As the dimension of the Jordan block increases, it becomes difficult to find the ``optimal canonical form'' for $N$.
  \item Theorem \ref{Bogoliubov SVD} also excludes the case where  $\Ker N$ is a strict subspace of $\Ker \mathcal{N}$ (it is always a subspace). In this case, $N$ is called $J$-degenerate. When $\Ker \mathcal{N}=\Ker N$, $N$ is called $J$-nondegenerate, and this is the generic situation for a doubled-up $N^{2m \times 2n}$, with $m \leq n$. The proof is as follows: We have that, $\Rank N^{\flat} =\Rank J_{2n} N^{\dag} J_{2m}= \Rank N^{\dag} = \Rank N$, because $J_{2k}$ is full rank for any $k$. From this follows that, $\Rank \mathcal{N}\leq \min (\Rank N, \Rank N^{\flat})= \Rank N$. Now, from Sylvester's rank inequality, we also have that $\Rank \mathcal{N} \geq \Rank N + \Rank N^{\flat}-2m = 2\,\Rank N -2m$. Hence, $2\,\Rank N -2m \leq \Rank \mathcal{N} \leq \Rank N$. Let us define the unitary $2k \times 2k$ matrix $\Phi_{2k}$ by $\Phi_{2k}=\frac{1}{\sqrt{2}}\bigl(\begin{smallmatrix}I_k & I_k \\ -\imath I_k & \imath I_k \end{smallmatrix}\bigr)$. Then, $N_{R}=\Phi_{2m}N\Phi_{2n}^{-1}$ is a real $2m \times 2n$ matrix. Conversely, given any real $2m \times 2n$ matrix $X$, we may create a $2m \times 2n$ doubled-up (complex) matrix $X_{D}$ by $X_{D}=\Phi_{2m}^{-1}X\Phi_{2n}$. Notice that $(X_D)_R=X$, and $(N_R)_D=N$, so there is an isomorphism between $2m \times 2n$ real matrices and $2m \times 2n$ complex doubled-up matrices. Also, $\Rank N_R=\Rank N \Leftrightarrow \Rank X_D=\Rank X$. It is a well known fact that a $p \times q$ real matrix, with $p \leq q$, will have rank equal to $p$, generically. A proof of this fact can be easily constructed by using the SVD and arguments from \cite[Section 5.6]{hirsmadev04}. Hence, it follows that the generic $2m \times 2n$ doubled-up matrix with $m \leq n$ has rank equal to $m$. For the corresponding $\mathcal{N}$, we have that $2(2m)-2m \leq \Rank \mathcal{N} \leq 2m \Rightarrow \Rank\mathcal{N}=2m$. Then, $\dim\Ker \mathcal{N}=2n-2m=\dim\Ker N$, from which $\Ker \mathcal{N}=\Ker N$ follows (recall that $\Ker N$ is a subspace of $\Ker \mathcal{N}$).  In the case $m \geq n$, one can similarly show that $\Ker N N^{\flat} =\Ker N^{\flat}$, generically. Then, one can prove Theorem \ref{Bogoliubov SVD} using $N^{\flat}$ in place of $N$. In Remark 2 after the proof of Theorem \ref{Bogoliubov SVD} in the Appendix, we demonstrate the fundamental issue with the $J$-degenerate case. Also, we identify a special situation where we can extend the validity of Theorem \ref{Bogoliubov SVD} in spite of $N$ being $J$-degenerate.
  \item We saw in the previous remark that there exists an isomorphism between complex doubled-up matrices and real matrices of the same dimensions. Indeed, given a $2m \times 2n$ complex doubled-up matrix $N$, $\Phi_{2m}N\Phi_{2n}^{-1}$ is a real $2m \times 2n$ matrix, where the unitary $2k \times 2k$ matrix $\Phi_{2k}$ is defined by $\Phi_{2k}= \frac{1}{\sqrt{2}}\bigl(\begin{smallmatrix}I_k & I_k \\ -\imath I_k & \imath I_k \end{smallmatrix}\bigr)$. Conversely, given any real $2m \times 2n$ matrix $X$, $\Phi_{2m}^{-1}X\Phi_{2n}$ is a $2m \times 2n$ doubled-up (complex) matrix. Define the symplectic unit matrix in $2k$ dimensions by $\mathbb{J}_{2k}= \bigl(\begin{smallmatrix} 0_{k \times k} & I_k \\ -I_k & 0_{k \times k} \end{smallmatrix}\bigr)$. Then, $\Phi_{2k} J_{2k} \Phi_{2k}^{-1}=\imath \mathbb{J}_{2k}$. For a $2m \times 2n$ real matrix $X$, define its $\sharp$-\emph{adjoint} $X^{\sharp}$, by $X^{\sharp}= -\mathbb{J}_{2n}X^{\top}\mathbb{J}_{2m}$. The $\sharp$-\emph{adjoint} satisfies properties similar to the usual adjoint, namely $(x_1 A + x_2 B)^{\sharp}=x_1 A^{\sharp} + x_2 B^{\sharp}$, and $(AB)^{\sharp}=B^{\sharp}  A^{\sharp}$. From the above definitions, it follows that $\Phi_{2n} N^{\flat} \Phi_{2m}^{-1}= -\mathbb{J}_{2n} (\Phi_{2m} N \Phi_{2n}^{-1})^{\dag} \mathbb{J}_{2m}=(\Phi_{2m} N \Phi_{2n}^{-1})^{\sharp}$. Then, given a $2k \times 2k$ Bogoliubov matrix $T$, $S=\Phi_{2k}T\Phi_{2k}^{-1}$ is a real matrix that satisfies $SS^{\sharp}=S^{\sharp} S=I_{2k}$, due to the fact that $S^{\sharp}=(\Phi_{2k}T\Phi_{2k}^{-1})^{\sharp}= \Phi_{2k}T^{\flat}\Phi_{2k}^{-1}= \Phi_{2k}T^{-1}\Phi_{2k}^{-1}=S^{-1}$. Such a matrix $S$ is called real symplectic. The set of these matrices forms a non-compact Lie group known as the real symplectic group  which is homomorphic to the Bogoliubov group. Using the previous definitions, Theorem \ref{Bogoliubov SVD} may be restated as follows:
\begin{theorem} \label{Symplectic SVD}
Let $X^{2m \times 2n}$ be a real matrix, and let $\mathcal{X} \doteq X^{\sharp}X$. We assume that all the eigenvalues of $\mathcal{X}$ are semisimple. Also, we assume that $\Ker \mathcal{X}=\Ker X$, i.e all the eigenvectors of $\mathcal{X}$ with zero eigenvalue belong to the kernel of $X$. Let $\lambda_i^+ >0, i=1,\ldots,r_+$, $\lambda_i^- <0, i=1,\ldots,r_-$, and $\lambda_i^c$, with $\Im\lambda_i^c >0$, $i=1\ldots,r_c$, be the eigenvalues of $\mathcal{X}$ that are, respectively, positive, negative, and non-real with positive imaginary part. Then, there exist symplectic matrices $V^{2m \times 2m}$, $W^{2n \times 2n}$, and a real matrix $\hat{X}^{2m \times 2n}=\bigl(\begin{smallmatrix} \hat{X}_1 & \hat{X}_2 \\ \hat{X}_2 & \hat{X}_3 \end{smallmatrix}\bigr)$, such that $X = V \, \hat{X} \, W^{\sharp}$, where $\hat{X}_{l}= \bigl(\begin{smallmatrix} \bar{X}_{l} & \mathbf{0} \\ \mathbf{0} & \mathbf{0} \end{smallmatrix}\bigr)$, for $l=1,2,3$, and
\begin{eqnarray*}
\bar{X}_1^{r \times r} &=& \diag(\sqrt{\lambda_1^+},\ldots,\sqrt{\lambda_{r_+}^+},\sqrt{|\lambda_1^-|},\ldots,\sqrt{|\lambda_{r_-}^-|},\alpha_1 I_2,\ldots,\alpha_{r_c} I_2), \\
\bar{X}_2^{r \times r} &=& \diag(\underbrace{0,\ldots,0}_{r_+ + r_-},\beta_1\mathbb{J}_2,\ldots,\beta_{r_c}\mathbb{J}_2), \\
\bar{X}_3^{r \times r} &=& \diag(\sqrt{\lambda_1^+},\ldots,\sqrt{\lambda_{r_+}^+},-\sqrt{|\lambda_1^-|},\ldots,-\sqrt{|\lambda_{r_-}^-|},\alpha_1 I_2,\ldots,\alpha_{r_c} I_2), \\
\end{eqnarray*}
where $r=r_+ + r_- + 2r_c$, and $\mathbb{J}_2 = \bigl(\begin{smallmatrix} 0 & 1 \\ -1 & 0 \end{smallmatrix} \bigr)$. The parameters $\alpha_i$ and $\beta_i$ are determined in terms of $\lambda_i^c$, as follows:
\[ \alpha_i = \sqrt{\frac{|\lambda_i^c|+\Re \lambda_i^c}{2}}, \ \ \beta_i = \frac{\Im \lambda_i^c}{\sqrt{2\big(|\lambda_i^c|+\Re \lambda_i^c\big)}}. \]
\end{theorem}
  \item Figures \ref{Passive_Network_1}, \ref{Passive_Network_2}, \ref{Passive_Network_3}, and \ref{Active_Network_1} may create the impression that, the $n-r$ (reduced system) modes that are not influenced directly by the inputs, are always controllable through the (obviously controllable) $r$ modes directly influenced by the inputs. However, this is not always the case. In the passive case, it is straightforward to see that if the unitary feedback gain $R$ is block-diagonal, $R= \bigl(\begin{smallmatrix} R_r & 0 \\ 0 & R_{n-r} \end{smallmatrix}\bigr)$, where $R_r$ and $R_{n-r}$ are unitary $r \times r$ and $(n-r) \times (n-r)$ matrices, respectively, then the $n-r$ modes that are not influenced directly by the inputs are uncontrollable (and unobservable). Moreover, it can be proven that this is the only mechanism through which the reduced and, equivalently, the original LQSS can lose controllability and observability. In the general case, the situation is more complicated. If we let $R= \bigl(\begin{smallmatrix} R_1 & R_2 \\ R_2^{\#} & R_1^{\#} \end{smallmatrix}\bigr)$, then $R_1$ and $R_2$ being block-diagonal (with blocks of dimensions $r$ and $n-r$), implies that the $n-r$ modes that are not influenced directly by the inputs are uncontrollable (and unobservable). However, this is not the only mechanism through which the reduced and, equivalently, the original LQSS can lose controllability or observability.

\end{enumerate}

\section*{Appendix}

This appendix contains the proof of Theorem \ref{Bogoliubov SVD}, along with some remarks. We begin with some definitions:
\begin{enumerate}
\item The ``sip'' matrix in $k$ dimensions is defined by the expression
\begin{equation*}
\mathcal{S}_k \doteq \left(\begin{array}{cccc} 0 & \cdots & 0 & 1 \\ 0 & \cdots & 1 & 0 \\ \vdots & \reflectbox{$\ddots$} & \vdots & \vdots \\ 1 & \cdots & 0 & 0 \\ \end{array}\right).
\end{equation*}
\item We define $\mathbf{j}_k(\lambda)$ to be the upper Jordan block of size $k$ with eigenvalue $\lambda$, if $\lambda$ is real, and the direct sum of two Jordan blocks of size $k/2$ each (for even $k$), the first with eigenvalue $\lambda$, and the second with eigenvalue $\lambda^*$, if $\lambda$ is complex. The matrix  whose columns are the eigenvector and the generalized eigenvectors corresponding to $\lambda$, in sequence  \cite{horjoh85}, will be called the eigenvector block corresponding to $\lambda$.
\item We define the matrix $\tilde{\Sigma}_{2k}= \bigl(\begin{smallmatrix} 0_{k \times k} & \imath I_k \\ -\imath I_k & 0_{k \times k} \end{smallmatrix}\bigr)$. Then,
we have that $\tilde{\Sigma}_{2k}^2=I_{2k}$, $\tilde{\Sigma}_{2k} \,\mathcal{S}_{2k}\,\tilde{\Sigma}_{2k}=-\mathcal{S}_{2k}$, and $\tilde{\Sigma}_{2k}\, \mathbf{j}_{2k}(\lambda)\,\tilde{\Sigma}_{2k}=\mathbf{j}_{2k}(\lambda^*)$. When its dimension can be inferred from context, it will be denoted simply by $\tilde{\Sigma}$.
\end{enumerate}

The $2n \times 2n$ matrix $\mathcal{N}\doteq N^{\flat}N$ is $\flat$-Hermitian, i.e. $\mathcal{N}^{\flat}=\mathcal{N}$. The spectral theorem for self-adjoint matrices in spaces with indefinite scalar products \cite{gohlanrod83} applied to the case of $\mathcal{N}$ as a $\flat$-Hermitian matrix in the Krein space $(\mathbb{C}^{2n}, J)$ takes the following form:
\begin{lemma}\label{Spectral decomposition of self-adjoint matrices in spaces with indefinite scalar products}
Let $\lambda_1$, \ldots, $\lambda_A$ be the real eigenvalues of $\mathcal{N}$, and $\lambda_{A+1}$, \ldots, $\lambda_{B}$ its complex eigenvalues. There exists a basis of $\mathbb{C}^{2n}$ in which the matrices $\mathcal{N}$ and $J$ have the following canonical forms:
\begin{eqnarray}
\mathcal{N}&=& \mathbf{j}_{k_1}(\lambda_1)\oplus\ldots\oplus\mathbf{j}_{k_A}(\lambda_A)\oplus\mathbf{j}_{k_{A+1}}(\lambda_{A+1})\oplus\ldots\oplus\mathbf{j}_{k_B}(\lambda_B),
\nonumber \\
J&=&\varepsilon_1\mathcal{S}_{k_1}\oplus\ldots\oplus\varepsilon_A\mathcal{S}_{k_A}\oplus\mathcal{S}_{k_{A+1}}\oplus\ldots\oplus\mathcal{S}_{k_B}, \label{Spectral thm for cal_N}
\end{eqnarray}
where $\varepsilon_i=\pm 1,\ i=1,\ldots,A$. This decomposition is unique except for permutations.
\end{lemma}
Let $\{z_1,\ldots,z_{2n}\}$ be the aforementioned basis of $\mathbb{C}^{2n}$. Let $Z^{2n \times 2n} \doteq [z_1 \ldots z_{2n}]$, and $Z_i^{2n \times k_i}$ the submatrix of $Z$ that contains the eigenvectors of the $i$-th block, for $i=1,\ldots,A,A+1,\ldots,B$. Then, (\ref{Spectral thm for cal_N}) can be expressed as follows:
\begin{eqnarray}
\mathcal{N}Z&=&Z\,\hat{\mathcal{N}}, \nonumber \\
J Z &=&Z\,\hat{J}, \label{Spectral thm for cal_N, version 2}
\end{eqnarray}
where the block-diagonal matrices $\hat{\mathcal{N}}$ and $\hat{J}$ are defined by the following expressions:
\begin{eqnarray}
\hat{\mathcal{N}}&\doteq& \diag(\mathbf{j}_{k_1}(\lambda_1),\ldots,\mathbf{j}_{k_A}(\lambda_A),\mathbf{j}_{k_{A+1}}(\lambda_{A+1}),\ldots,\mathbf{j}_{k_B}(\lambda_B)), \nonumber\\
\hat{J} &\doteq& \diag(\varepsilon_1\mathcal{S}_{k_1},\ldots,\varepsilon_A\mathcal{S}_{k_A},\mathcal{S}_{k_{A+1}},\ldots,\mathcal{S}_{k_B}). \label{cal_N_hat and J_hat}
\end{eqnarray}
Furthermore, if we define $\varepsilon_i \doteq 1$ for $i=A+1,\ldots,B$, (\ref{Spectral thm for cal_N}) implies that $Z_i^{\dag}\, J \, Z_j = \delta_{ij}\, (\varepsilon_i \mathcal{S}_i)$, for $i,j=1,\ldots,A,A+1,\ldots,B$. That is, the different blocks appearing in (\ref{Spectral thm for cal_N}) are $J$-orthogonal.

Besides being $\flat$-Hermitian, $\mathcal{N}$ is also doubled-up, i.e. $\Sigma \mathcal{N}\Sigma=\mathcal{N}^{\#}$. From the first equation of (\ref{Spectral thm for cal_N, version 2}), we compute:
\begin{eqnarray}
&&(\Sigma \mathcal{N}\Sigma)\,(\Sigma Z) = (\Sigma Z)\,\hat{\mathcal{N}} \nonumber \\
&\Rightarrow& \mathcal{N}^{\#}(\Sigma Z) = (\Sigma Z)\, \hat{\mathcal{N}} \nonumber \\
&\Rightarrow& \mathcal{N}(\Sigma Z^{\#}) = (\Sigma Z^{\#})\, \hat{\mathcal{N}}^{\#}. \label{consequence 1 of doubled-up_cal_N}
\end{eqnarray}
Similarly, from the second equation of (\ref{Spectral thm for cal_N, version 2}), we have:
\begin{eqnarray}
&&(\Sigma J \Sigma)\,(\Sigma Z) = (\Sigma Z)\,\hat{J} \nonumber \\
&\Rightarrow& -J\,(\Sigma Z) = (\Sigma Z)\, \hat{J} \nonumber \\
&\Rightarrow& J\,(\Sigma Z^{\#}) = (\Sigma Z^{\#})\, (-\hat{J}). \label{consequence 2 of doubled-up_cal_N}
\end{eqnarray}
If we restrict (\ref{consequence 1 of doubled-up_cal_N}) and (\ref{consequence 2 of doubled-up_cal_N}) in the real eigenspace of $\mathcal{N}$, $Z_r \doteq [Z_{k_1} \ldots Z_{k_A}]$, we obtain the following:
\begin{eqnarray*}
\mathcal{N}\, (\Sigma Z_r^{\#})&=&(\Sigma Z_r^{\#})\, \diag(\mathbf{j}_{k_1}(\lambda_1),\ldots,\mathbf{j}_{k_A}(\lambda_A)), \\
J\, (\Sigma Z_r^{\#})&=&(\Sigma Z_r^{\#})\, \diag(-\varepsilon_1\mathcal{S}_{k_1},\ldots,-\varepsilon_A\mathcal{S}_{k_A}).
\end{eqnarray*}
The uniqueness of the decomposition (\ref{Spectral thm for cal_N, version 2}) implies that for every real eigenvalue $\lambda$, there are two eigenvector blocks, say $Z_i$ and $Z_j$, such that $Z_j = \Sigma\, Z_i^{\#}$, and $\varepsilon_j=-\varepsilon_i$. The situation for the complex eigenvalues is a bit more complicated. The restriction of equations (\ref{consequence 1 of doubled-up_cal_N}) and (\ref{consequence 2 of doubled-up_cal_N}) in the complex eigenspace of $\mathcal{N}$, $Z_c \doteq [Z_{k_{A+1}} \ldots Z_{k_B}]$, furnishes the following relations:
\begin{eqnarray*}
\mathcal{N}\, (\Sigma Z_c^{\#})&=&(\Sigma Z_c^{\#})\, \diag(\mathbf{j}_{k_{A+1}}(\lambda_{A+1}^*),\ldots,\mathbf{j}_{k_B}(\lambda_B^*)), \\
J\, (\Sigma Z_c^{\#})&=&(\Sigma Z_c^{\#})\, \diag(-\mathcal{S}_{k_{A+1}},\ldots,-\mathcal{S}_{k_B}).
\end{eqnarray*}
By defining the matrix $\tilde{\Sigma}_c \doteq \diag(\tilde{\Sigma}_{k_{A+1}},\ldots,\tilde{\Sigma}_{k_B})$, the equations above can be rewritten as follows:
\begin{eqnarray*}
\mathcal{N}\, (\Sigma Z_c^{\#})&=&(\Sigma Z_c^{\#})\, \tilde{\Sigma}_c\, \diag(\mathbf{j}_{k_{A+1}}(\lambda_{A+1}),\ldots,\mathbf{j}_{k_B}(\lambda_B))\, \tilde{\Sigma}_c, \\
J\, (\Sigma Z_c^{\#})&=&(\Sigma Z_c^{\#})\, \tilde{\Sigma}_c\, \diag(\mathcal{S}_{k_{A+1}},\ldots,\mathcal{S}_{k_B})\, \tilde{\Sigma}_c.
\end{eqnarray*}
Multiplying both equations from the right with $\tilde{\Sigma}_c$, provides the desired form:
\begin{eqnarray*}
\mathcal{N}\, (\Sigma Z_c^{\#}\tilde{\Sigma}_c)&=&(\Sigma Z_c^{\#}\tilde{\Sigma}_c)\, \diag(\mathbf{j}_{k_{A+1}}(\lambda_{A+1}),\ldots,\mathbf{j}_{k_B}(\lambda_B)), \\
J\, (\Sigma Z_c^{\#}\tilde{\Sigma}_c)&=&(\Sigma Z_c^{\#}\tilde{\Sigma}_c)\, \diag(\mathcal{S}_{k_{A+1}},\ldots,\mathcal{S}_{k_B}).
\end{eqnarray*}
Invoking the uniqueness of the decomposition (\ref{Spectral thm for cal_N, version 2}) again, implies that for every complex eigenvalue $\lambda$, there are two eigenvector blocks, say $Z_i$ and $Z_j$, such that $Z_j = \Sigma\, Z_i^{\#}\tilde{\Sigma}_{k_i}$. Now we are ready to prove Theorem \ref{Bogoliubov SVD}.\\

\textbf{Proof of Theorem \ref{Bogoliubov SVD}:} We begin with the real positive eigenvalues, $\lambda^+_i, i=1,\ldots,r_+$. To each one there correspond two eigenvectors, $z_i^+$ with $(z_i^+)^{\dag} J z_i^+ =1$, and $\Sigma z_i^{+\#}$ with $(\Sigma z_i^{+\#})^{\dag} J (\Sigma z_i^{+\#}) =-1$ (we adopt the convention of expressing the eigenvector whose inner product with itself is negative in terms of the eigenvector whose inner product with itself is positive). These two eigenvectors are also $J$-orthogonal to each other, i.e $(z_i^+)^{\dag} J (\Sigma z_i^{+\#}) =0$. Due to the semi-simplicity hypothesis and the uniqueness of the decomposition (\ref{Spectral thm for cal_N, version 2}), different eigenspaces are $J$-orthogonal to each other, as well, so that
\begin{eqnarray*}
z_i^{+\dag}J\, z_j^+ &=& 0, \\
(\Sigma z_i^{+\#})^\dag J\, (\Sigma z_j^{+\#}) &=& 0, \\
(\Sigma z_i^{+\#})^\dag J\, z_j^+ &=& 0,
\end{eqnarray*}
for $i \neq j =1,\ldots,r_+$. If we define the $2n \times r_+$ matrix $Z^+ \doteq [z_1^+ \, \ldots\, z_{r_+}^+]$, it is straightforward to see that
\[ [Z^+ \, \Sigma Z^{+\#}]^{\dag} \, J \, [Z^+ \, \Sigma Z^{+\#}] = J_{2r_+}, \]
and
\[ \mathcal{N}\, [Z^+ \, \Sigma Z^{+\#}]= \, [Z^+ \, \Sigma Z^{+\#}]\, \diag(\lambda_1^+,\ldots,\lambda_{r_+}^+,\lambda_1^+,\ldots,\lambda_{r_+}^+). \]
The treatment of the real negative eigenvalues is identical. The resulting $2n \times r_-$ matrix $Z^-$ satisfies the analogous relations
\[ [Z^- \, \Sigma Z^{-\#}]^{\dag} \, J \, [Z^- \, \Sigma Z^{-\#}] = J_{2r_-}, \]
and
\[ \mathcal{N}\, [Z^- \, \Sigma Z^{-\#}]= \, [Z^- \, \Sigma Z^{-\#}] \,\diag(\lambda_1^-,\ldots,\lambda_{r_-}^-,\lambda_1^-,\ldots,\lambda_{r_-}^-). \]
Similarly, for the case of zero eigenvalues the corresponding $2n \times r_0$ matrix $Z^0$ ($r_0$ is the number of zero eigenvalues) satisfies the relations
\begin{equation*}
[Z^0 \, \Sigma Z^{0\#}]^{\dag} \, J \, [Z^0 \, \Sigma Z^{0\#}] = J_{2r_0},
\end{equation*}
and
\begin{eqnarray*}
\mathcal{N}\, [Z^0 \, \Sigma Z^{0\#}] &=& 0_{2n \times 2r_0}, \\
N \, [Z^0 \, \Sigma Z^{0\#}] &=& 0_{2m \times 2r_0}.
\end{eqnarray*}
Let $\lambda_i^c=\mu_i + \imath\, \nu_i$, with $\nu_i>0$, $i=1\ldots,r_c$, denote the non-real eigenvalues of $\mathcal{N}$ with positive imaginary part. To each one, there correspond four associated eigenvectors, $z_{i1}^c$, $z_{i2}^c$, $\Sigma z_{i2}^{c\#}$, and $\Sigma z_{i1}^{c\#}$, where
\begin{eqnarray*}
\mathcal{N}\, z_{i1}^c & = & \lambda_i^c \, z_{i1}^c, \\
\mathcal{N}\, z_{i2}^c & = & \lambda_i^{c*} \, z_{i2}^c, \\
\mathcal{N}\, (\Sigma z_{i2}^{c\#}) & = & \lambda_i^c \, (\Sigma z_{i2}^{c\#}), \\
\mathcal{N}\, (\Sigma z_{i1}^{c\#}) & = & \lambda_i^{c*} \, (\Sigma z_{i1}^{c\#}),
\end{eqnarray*}
and
\begin{eqnarray*}
z_{i\alpha}^{c\dag}\, J \, z_{i\beta}^c & = & 1-\delta_{\alpha \beta}, \\
(\Sigma z_{i\alpha}^{c\#})^{\dag} \, J \, (\Sigma z_{i\beta}^{c\#}) & = & -1+\delta_{\alpha \beta}, \\
(\Sigma z_{i\alpha}^{c\#})^{\dag} \, J \, z_{i\beta}^c &=& 0, \ \mathrm{for} \ \alpha,\beta=1,2.
\end{eqnarray*}
For our purposes, it will be beneficial to work with the following linear combinations:
\begin{eqnarray*}
\tilde{z}_{i1}^c &\doteq& \frac{1}{\sqrt{2}}\,(z_{i1}^c + z_{i2}^c),\\
\tilde{z}_{i2}^c &\doteq& \frac{1}{\sqrt{2}}\,(z_{i1}^c - z_{i2}^c),
\end{eqnarray*}
along with $\Sigma \tilde{z}_{i1}^{c\#}$, and $\Sigma \tilde{z}_{i2}^{c\#}$. It is straightforward to show that
\begin{eqnarray*}
\tilde{z}_{i1}^{c\dag}\, J \, \tilde{z}_{i1}^c  & = & (\Sigma \tilde{z}_{i2}^{c\#})^{\dag}\, J \, (\Sigma \tilde{z}_{i2}^{c\#})= 1,\\
\tilde{z}_{i2}^{c\dag}\, J \, \tilde{z}_{i2}^c  & = & (\Sigma \tilde{z}_{i1}^{c\#})^{\dag}\, J \, (\Sigma \tilde{z}_{i1}^{c\#})= -1,\\
\tilde{z}_{i1}^{c\dag}\, J \, \tilde{z}_{i2}^c  & = & (\Sigma \tilde{z}_{i1}^{c\#})^{\dag}\, J \, (\Sigma \tilde{z}_{i2}^{c\#})= 0, \\
(\Sigma \tilde{z}_{i\alpha}^{c\#})^{\dag}\, J \, \tilde{z}_{i\beta}^c  & = & 0, \ \mathrm{for} \ \alpha,\beta=1,2.
\end{eqnarray*}
and
\begin{eqnarray*}
&& \mathcal{N}\, [\tilde{z}_{i1}^c \, \Sigma \tilde{z}_{i2}^{c\#}\, \Sigma \tilde{z}_{i1}^{c\#}\, \tilde{z}_{i2}^c] \\
&=& [\tilde{z}_{i1}^c \, \Sigma \tilde{z}_{i2}^{c\#}\, \Sigma \tilde{z}_{i1}^{c\#}\, \tilde{z}_{i2}^c]\,
\left(\begin{array}{cc|cc}
\mu_i & 0 & 0 & \imath\,\nu_i \\
0 & \mu_i & -\imath\,\nu_i & 0 \\ \hline
0 & -\imath\,\nu_i & \mu_i & 0 \\
\imath\,\nu_i & 0 & 0 & \mu_i \\
\end{array}\right) \\
&=& [\tilde{z}_{i1}^c \, \Sigma \tilde{z}_{i2}^{c\#}\, \Sigma \tilde{z}_{i1}^{c\#}\, \tilde{z}_{i2}^c]\,
\left(\begin{array}{cc} \mu_i\, I_2 & -\nu_i\, \sigma_2 \\ \nu_i\, \sigma_2 & \mu_i\, I_2 \\ \end{array}\right),
\end{eqnarray*}
where $\sigma_2 = \bigl(\begin{smallmatrix} 0 & -\imath \\ \imath & 0 \end{smallmatrix} \bigr)$ is one of the Pauli matrices.
Hence, if we define $Z^c \doteq [\tilde{z}_{11}^c \, \Sigma \tilde{z}_{12}^{c\#}\, \ldots \, \tilde{z}_{r_c 1}^c \, \Sigma \tilde{z}_{r_c 2}^{c\#}]$, and recall that different eigenvalue blocks are $J$-orthogonal to each other, we can see that the following relations hold:
\[[Z^c \, \Sigma Z^{c\#}]^{\dag} \,J\, [Z^c \, \Sigma Z^{c\#}] = J_{4r_c}, \]
and
\[ \mathcal{N}\, [Z^c \, \Sigma Z^{c\#}]= \, [Z^c \, \Sigma Z^{c\#}]\,
\left(\begin{array}{ccc|ccc}
\mu_1\, I_2 &  &  & -\nu_1\, \sigma_2 &  &  \\
 & \ddots &  &  & \ddots  &  \\
 &  & \mu_{r_c}\, I_2 &  &  & -\nu_{r_c}\, \sigma_2 \\ \hline
\nu_1\, \sigma_2 &  &  & \mu_1\, I_2 &  &  \\
 & \ddots &  &  & \ddots &  \\
 &  & \nu_{r_c}\, \sigma_2 &  &  & \mu_{r_c}\, I_2 \\
\end{array}\right). \]
To put the various cases together, we define
\[ W \doteq \big[\, [Z^+\, Z^-\, Z^c\, Z^0] \ \Sigma [Z^+\, Z^-\, Z^c\, Z^0]^{\#} \,\big]. \]
This $2n\times 2n$ matrix is Bogoliubov. Indeed, recalling the orthonormality relations within each case (complex, real positive, real negative, and zero eigenvalues), and the fact that different case blocks are $J$-orthogonal to each other, we can see that
\[ W^{\flat}W=J(W^{\dag}JW)=JJ=I,  \]
and,
\begin{eqnarray*}
\Sigma W \Sigma &=& \Sigma \big[\, \Sigma [Z^+\, Z^-\, Z^c\, Z^0]^{\#}\ [Z^+\, Z^-\, Z^c\, Z^0] \,\big] \\
&=& \big[\, [Z^+\, Z^-\, Z^c\, Z^0]^{\#} \ \Sigma [Z^+\, Z^-\, Z^c\, Z^0] \,\big] \\
&=& W^{\#}.
\end{eqnarray*}
We also have that,
\begin{eqnarray}
&& \mathcal{N}\,\big[\, [Z^+\, Z^-\, Z^c] \ \Sigma [Z^+\, Z^-\, Z^c]^{\#} \,\big] =
\big[\, [Z^+\, Z^-\, Z^c] \ \Sigma [Z^+\, Z^-\, Z^c]^{\#} \,\big]\,\bar{\mathcal{N}}, \label{cal_N in subspace orthogonal to its kernel}
\end{eqnarray}
where
\[\bar{\mathcal{N}} \doteq \left(\begin{array}{cc} \bar{\mathcal{N}}_1 & \bar{\mathcal{N}}_2 \\ \bar{\mathcal{N}}_2^{\#} & \bar{\mathcal{N}}_1^{\#} \\ \end{array}\right),\]
and
\begin{eqnarray*}
\bar{\mathcal{N}}_1 &=& \diag(\lambda_1^+,\ldots,\lambda_{r_+}^+,\lambda_1^-,\ldots,\lambda_{r_-}^-,\mu_1\, I_2,\ldots,\mu_{r_c}\, I_2,), \\
\bar{\mathcal{N}}_2 &=& \diag(\underbrace{0,\ldots,0}_{r_+ +\, r_-},-\nu_1\, \sigma_2,\ldots,-\nu_{r_c}\, \sigma_2).
\end{eqnarray*}
$\bar{\mathcal{N}}$ is just the restriction of $\mathcal{N}$ on its $r$-dimensional invariant subspace spanned by eigenvectors with non-trivial eigenvalues ($r=r_+ + r_- +2r_c$). We can factor $\bar{\mathcal{N}}=\bar{N}^{\flat}\bar{N}$ with $\bar{N} \doteq \bigl(\begin{smallmatrix} \bar{N}_1 & \bar{N}_2 \\ \bar{N}_2^{\#} & \bar{N}_1^{\#} \end{smallmatrix}\bigr)$, where
\begin{eqnarray*}
\bar{N}_1 &=& \diag(\sqrt{\lambda_1^+},\ldots,\sqrt{\lambda_{r_+}^+},\underbrace{0,\ldots,0}_{r_-},\alpha_1 I_2,\ldots,\alpha_{r_c} I_2), \\
\bar{N}_2 &=& \diag(\underbrace{0,\ldots,0}_{r_+},\sqrt{|\lambda_1^-|},\ldots,\sqrt{|\lambda_{r_-}^-|},-\beta_1\sigma_2,\ldots,-\beta_{r_c}\sigma_2).
\end{eqnarray*}
The parameters $\alpha_i$ and $\beta_i$ are determined in terms of $\lambda_i^c$, as follows:
\[ \alpha_i = \sqrt{\frac{|\lambda_i^c|+\Re \lambda_i^c}{2}}, \ \ \beta_i = \frac{\Im \lambda_i^c}{\sqrt{2\big(|\lambda_i^c|+\Re \lambda_i^c\big)}}. \]
Introducing the definition $\mathcal{N}= N^{\flat}N$, and the factorization $\bar{\mathcal{N}}=\bar{N}^{\flat}\bar{N}$ into (\ref{cal_N in subspace orthogonal to its kernel}), we compute:
\begin{eqnarray*}
&&N^{\flat}N \big[\, [Z^+\, Z^-\, Z^c] \ \Sigma [Z^+\, Z^-\, Z^c]^{\#} \,\big] \\
&=& \big[\, [Z^+\, Z^-\, Z^c] \ \Sigma [Z^+\, Z^-\, Z^c]^{\#} \,\big]\, \bar{N}^{\flat}\bar{N}\, \Rightarrow \\
&&\big[\, [Z^+\, Z^-\, Z^c] \ \Sigma [Z^+\, Z^-\, Z^c]^{\#} \,\big]^{\flat}\,N^{\flat} N\big[\, [Z^+\, Z^-\, Z^c] \ \Sigma [Z^+\, Z^-\, Z^c]^{\#} \,\big] \\
&=& \bar{N}^{\flat}\bar{N}\, \Rightarrow \\
&& \Big( N\big[\, [Z^+\, Z^-\, Z^c] \ \Sigma [Z^+\, Z^-\, Z^c]^{\#} \,\big]\,(\bar{N})^{-1}\Big)^{\flat} \cdot \\
&& \Big( N\big[\, [Z^+\, Z^-\, Z^c] \ \Sigma [Z^+\, Z^-\, Z^c]^{\#} \,\big]\,(\bar{N})^{-1}\Big) = I.
\end{eqnarray*}
The fact that $\bar{N}$ is a full rank square matrix of dimension $2r$ was implicitly used in the above calculation to guarrantee its invertibility. The $2m \times 2r$ matrix $N\big[\, [Z^+\, Z^-\, Z^c] \ \Sigma [Z^+\, Z^-\, Z^c]^{\#} \,\big]\,(\bar{N})^{-1}$ is doubled-up, since each of its factors has this property. Then, there exists a $2m \times r$ matrix $V_I$, such that
\begin{eqnarray}
N\big[\, [Z^+\, Z^-\, Z^c] \ \Sigma [Z^+\, Z^-\, Z^c]^{\#} \,\big]\,(\bar{N})^{-1} &=& [V_I \ \Sigma V_I^{\#}] \Leftrightarrow \nonumber \\
N\big[\, [Z^+\, Z^-\, Z^c] \ \Sigma [Z^+\, Z^-\, Z^c]^{\#} \,\big] &=& [V_I \ \Sigma V_I^{\#}]\, \bar{N}. \label{first half of Bogoliubov SVD}
\end{eqnarray}
Notice that the columns of $[V_I \ \Sigma V_I^{\#}]$ are $J$-orthonormal, i.e. $[V_I \ \Sigma V_I^{\#}]^{\dag}\, J \;[V_I \ \Sigma V_I^{\#}]=J_{2r}$. The final step is to complete a $J$-orthonormal basis of $(\mathbb{C}^{2m},J_{2m})$ with the doubled-up property, that is find a matrix $V_{II}^{2m \times (m-r)}$, such that
$V \doteq \big[\,[V_I \, V_{II}]\ \Sigma [V_I \, V_{II}]^{\#} \, \big] $ is Bogoliubov. To do this, consider the image of $[V_I \ \Sigma V_I^{\#}]$. It is a \emph{nondegenerate} subspace of $\mathbb{C}^{2m}$, meaning that it admits a $J$-orthonormal basis. Such a basis is in fact furnished by the columns of $[V_I \ \Sigma V_I^{\#}]$. It follows then \cite{gohlanrod83}, that its $J$-orthogonal complement in $\mathbb{C}^{2m}$, is also nondegenerate, hence it also admits a $J$-orthonormal basis. Any such basis must contain $m-r$ vectors whose inner product with themselves is 1, and as many whose inner product with themselves is -1. Then, $V_{II}$ can be any matrix whose columns are comprised by those basis vectors whose inner product with themselves is 1. Finally, combining equation (\ref{first half of Bogoliubov SVD}) along with
\[ N \, [Z^{0} \, \Sigma Z^{0\#}]=0_{2m \times 2r_0}=[\bar{V}_{II} \, \Sigma \bar{V}_{II}]\, 0_{2(m-r) \times 2r_0}, \]
we obtain the equation $N\,W=V\,\hat{N}$, where $\hat{N}$ has exactly the form in the statement of the theorem. Given that $W$ is Bogoliubov, the statement of the theorem follows.$\blacksquare$

We conclude this appendix with two remarks that extend the theorem in some special cases.\\[.25em]
\noindent \textbf{Remark 1}. Here, we extend Theorem \ref{Bogoliubov SVD} in the case of a real eigenvalue with a Jordan block of size $2$. We begin with some simple facts. It is easy to see that
\[ \mathcal{S}_{2k}=\left(\begin{array}{cc} 0_k & \mathcal{S}_k \\ \mathcal{S}_k & 0_k \\ \end{array}\right),\ \mathrm{and} \
\mathcal{S}_{2k+1}=\left(\begin{array}{lll} 0_k & 0_{k\times 1} & \mathcal{S}_k \\ 0_{1\times k} & 1 & 0_{1\times k} \\ \mathcal{S}_k & 0_{k\times 1} & 0_k \\ \end{array}\right). \]
From this structure, and the fact that $\mathcal{S}_k^2=I_k$, it can be proven easily that the matrices that diagonalize $\mathcal{S}_{2k}$ and $\mathcal{S}_{2k+1}$ are, respectively,
\[ T_{2k} \doteq \frac{1}{\sqrt{2}} \left(\begin{array}{cr} I_k & \mathcal{S}_k \\ \mathcal{S}_k & -I_k \\ \end{array}\right),\ \mathrm{and} \
T_{2k+1} \doteq \frac{1}{\sqrt{2}} \left(\begin{array}{lcr} I_k & 0_{k\times 1} & \mathcal{S}_k \\ 0_{1\times k} & \sqrt{2} & 0_{1\times k} \\ \mathcal{S}_k & 0_{k\times 1} & -I_k \\ \end{array}\right), \]
with $T_{2k}^{-1} \mathcal{S}_{2k} T_{2k}=J_{2k}$, and $T_{2k+1}^{-1} \mathcal{S}_{2k+1} T_{2k+1}=\diag(I_{k+1},-I_k)$.

Let us consider now the case of a real eigenvalue with a Jordan block of size $2$. Lemma \ref{Spectral decomposition of self-adjoint matrices in spaces with indefinite scalar products}, along with the discussion that follows it, implies the existence of two vectors, $z_1$ and $z_2$, such that, for $Z \doteq [z_1 \, z_2 \, \Sigma_4 z_1^{\#} \, \Sigma_4 z_2^{\#}]$, we have
\begin{eqnarray*}
J_4 Z &=& Z \left(\begin{array}{cr} \mathcal{S}_2 & 0_2 \\ 0_2 & -\mathcal{S}_2 \\ \end{array}\right), \ \mathrm{and} \\
\mathcal{N}Z &=& Z \left(\begin{array}{cc} \mathbf{j}_2(\lambda) & 0_2 \\ 0_2 & \mathbf{j}_2(\lambda) \\ \end{array}\right).
\end{eqnarray*}
The vectors $\bar{z}_1$ and $\bar{z}_2$ defined by
\begin{eqnarray*}
[\bar{z}_1 \, \bar{z}_2] \doteq [\frac{z_1 + z_2}{\sqrt{2}}\,\frac{z_1 - z_2}{\sqrt{2}}] = [z_1 \, z_2]\, T_2,
\end{eqnarray*}
satisfy the relation
\begin{eqnarray*}
J_4 [\bar{z}_1 \, \bar{z}_2] = [\bar{z}_1 \, \bar{z}_2] \,\big( T_2^{-1}\mathcal{S}_{2}T_2\big) = [\bar{z}_1 \, \bar{z}_2] J_2 .
\end{eqnarray*}
This means that $\bar{z}_1$ and $\bar{z}_2$ are $J_4$-orthonormal (with respective $J_4$-norms $\pm1$). We can construct the Bogoliubov matrix $W$ of Theorem \ref{Bogoliubov SVD} out of them, as $W= [\bar{z}_1 \, \Sigma_4 \bar{z}_2^{\#} \, \Sigma_4 \bar{z}_1^{\#} \bar{z}_2]= Z \mathcal{T}_4$, where
\begin{eqnarray*}
\mathcal{T}_4 \doteq \frac{1}{\sqrt{2}} \left(\begin{array}{crcr} 1&0&0&1 \\ 1&0&0&-1\\ 0&1&1&0\\ 0&-1&1&0 \\ \end{array}\right).
\end{eqnarray*}
The structure of $\mathcal{T}_4$ is inherited from that of $T_2$. We also have $\mathcal{N}W = W \bar{\mathcal{N}}$, where
\begin{eqnarray*}
\bar{\mathcal{N}} \doteq \mathcal{T}_4^{-1} \left(\begin{array}{cc} \mathbf{j}_2(\lambda) & 0_2 \\ 0_2 & \mathbf{j}_2(\lambda) \\ \end{array}\right) \mathcal{T}_4
= \left(\begin{array}{cc|cc}
\lambda + \frac{1}{2}&0&0&-\frac{1}{2} \\
0&\lambda - \frac{1}{2}&\frac{1}{2}&0\\
\hline 0&-\frac{1}{2}&\lambda + \frac{1}{2}&0\\
\frac{1}{2} &0&0&\lambda - \frac{1}{2} \\
\end{array}\right).
\end{eqnarray*}
To proceed, we have to factorize $\bar{\mathcal{N}}=\bar{N}^{\flat}\bar{N}$.  One such solution is given by
\begin{eqnarray*}
\bar{N}=\left(\begin{array}{cccc}
c \cosh x& \sinh x & 0 & -c \sinh x \\
0& c \cosh x & c \sinh x & \cosh x \\
0& -c \sinh x & c \cosh x & \sinh x \\
c \sinh x& \cosh x & 0 & c \cosh x \\
\end{array}\hspace{-.75em}\right),
\end{eqnarray*}
with $c=\sqrt{\lambda + \frac{1}{2}}$, and $\sinh 2x=\frac{1}{2c^2}$, for $\lambda \geq -\frac{1}{2}$, and
\begin{eqnarray*}
\bar{N}=\left(\begin{array}{cccc}
\cosh x & c \sinh x & c \cosh x & 0 \\
-c \sinh x& 0 & \sinh x & c \cosh x \\
c \cosh x& 0 & \cosh x & c \sinh x \\
\sinh x& c \cosh x & -c \sinh x & 0 \\
\end{array}\hspace{-.75em}\right),
\end{eqnarray*}
with $c=\sqrt{|\lambda - \frac{1}{2}|}$, and $\sinh 2x=-\frac{1}{2c^2}$, for $\lambda \leq \frac{1}{2}$. In both cases, the kernel of $\bar{N}$ is trivial for $\lambda=0$. Hence, for $\lambda=0$, this $\bar{N}$ is appropriate to use in the construction of $\hat{N}$ (see proof of Theorem \ref{Bogoliubov SVD}) only when $0$ is an eigenvalue of $\mathcal{N}$ whose eigenvector is not in $\Ker N$. For the case when $0$ is an eigenvalue of $\mathcal{N}$ whose eigenvector is in $\Ker N$, the following $\bar{N}$ is appropriate:
\begin{eqnarray*}
\bar{N}= \frac{1}{\sqrt{2}} \left(\begin{array}{rrrr}
1&0&0 &-1\\
0&0&0&0\\
0&-1&1&0\\
0&0&0&0\\
\end{array}\right).
\end{eqnarray*}
In every case, we can construct the Bogoliubov matrix $V$ of Theorem \ref{Bogoliubov SVD} following the steps of its proof. \vspace*{.25em}

\noindent \textbf{Remark 2.} In Theorem \ref{Bogoliubov SVD}, we required that $\Ker \mathcal{N}=\Ker N$. In this case, $N$ is called $J$-nondegenerate. This condition can be checked simply by calculating the rank of the matrices $N$ and $\mathcal{N}$. In general, $\Rank(\mathcal{N}) \leq \Rank(N)$, but when the two are equal, $N$ is $J$-nondegenerate. To describe the issue with $J$-degenerate matrices, we need some simple definitions and facts. Let $2r_{0*}$ be the number of (semisimple) zero eigenvalues of $\mathcal{N}$ whose corresponding eigenvectors are not in $\Ker N$ ($r_{0*} \leq n$). Let $z_i^{0*}, i=1,\ldots,r_{0*}$ be the corresponding $J$-orthonormal eigenvectors whose inner product with themselves is 1. Define $Z^{0*} \doteq [z_1^{0*}\, \ldots \, z_{r_{0*}}^{0*}]$, and $P \doteq N\, [Z^{0*}\, (\Sigma Z^{0*})^{\#}]$. In order to put $N$ in a canonical form, one should be able to write
\begin{equation}\label{Bogoliubov SVD, degenerate case}
P=N\, [Z^{0*}\, (\Sigma Z^{0*})^{\#}]=[V_{III} \ \Sigma V_{III}^{\#}]\, \bar{N}_{0*},
\end{equation}
where $[V_{III} \ \Sigma V_{III}^{\#}]$ would be the $J$-orthonormal basis of $\Ima(P)$, and $\bar{N}_{0*}$ the restriction/``reduced form'' of $N$ in that subspace. Then, one would use $Z^{0*}$ and $V_{III}$ in the construction of the Bogoliubov matrices $W$ and $V$, respectively, and $\bar{N}_{0*}$ in the construction of $\hat{N}$, in the proof of Theorem \ref{Bogoliubov SVD}. The problem is that
\[ P^{\flat}P= [Z^{0*}\, (\Sigma Z^{0*})^{\#}]^{\flat} \, \mathcal{N}\, [Z^{0*}\, (\Sigma Z^{0*})^{\#}] =\mathbf{0}.  \]
Thus, the columns of $P$ are a set of self and mutually $J$-orthogonal vectors. Hence, $\Ima P$ is a degenerate $2r_{0*}$-dimensional subspace of $\mathbb{C}^{2m}$, and degenerate subspaces do not have $J$-orthonormal bases. So, in the case $N$ is degenerate, the existence of a $J$-orthonormal basis for $\Ima P$ is forbidden.

In the following, we identify a special case in which it is possible to establish a relation analogous to (\ref{Bogoliubov SVD, degenerate case}), and use it to extend the applicability of Theorem \ref{Bogoliubov SVD} to the degenerate case. This is the case when an additional condition holds, namely $P P^{\flat} = \mathbf{0}$. Recall that $P$ is doubled-up because it is the product of two doubled-up matrices, and let $P=\bigl(\begin{smallmatrix} P_1 & P_2 \\ P_2^{\#} & P_1^{\#} \end{smallmatrix}\bigr)$. Equations $P^{\flat}P=0$, and $P\, P^{\flat}=0$, imply that $P_1^{\dag}P_1 = P_2^{\top} P_2^{\#}$, and $P_1\,P_1^{\dag} =P_2\,P_2^{\dag}$. Then, if $P_1=UHY^{\dag}$ is a SVD for $P_1$, it is straightforward to show that $P_2=UEHY^{\top}$, where $E=\diag(\pm1,\ldots,\pm1)$. Thus, we can factorize $P$ as follows:
\[ P= \left(\begin{array}{ll}
U & \mathbf{0} \\ \mathbf{0} & U^{\#} \\
\end{array}\right)
\left(\begin{array}{rr}
H & EH \\ EH & H \\ \end{array}\right)
\left(\begin{array}{ll}
Y^{\dag} & \mathbf{0} \\ \mathbf{0} & Y^{\top} \\
\end{array}\right).\]
Combined with the definition of $P$, the above equation leads to
\begin{eqnarray*}
&& N\, [Z^{0*}\, (\Sigma Z^{0*})^{\#}]= \left(\begin{array}{ll}
U & \mathbf{0} \\ \mathbf{0} & U^{\#} \\
\end{array}\right)
\left(\begin{array}{rr}
H & EH \\ EH & H \\ \end{array}\right)
\left(\begin{array}{ll}
Y^{\dag} & \mathbf{0} \\ \mathbf{0} & Y^{\top} \\
\end{array}\right) \\
&\Leftrightarrow& N\, [Z^{0*}\, (\Sigma Z^{0*})^{\#}] \left(\begin{array}{ll}
Y & \mathbf{0} \\ \mathbf{0} & Y^{\#} \\
\end{array}\right) =  \left(\begin{array}{ll}
U & \mathbf{0} \\ \mathbf{0} & U^{\#} \\
\end{array}\right)
\left(\begin{array}{rr}
H & EH \\ EH & H \\
\end{array}\right)  \\
&\Leftrightarrow& N\, [Z^{0*}Y\, (\Sigma Z^{0*})^{\#}Y^{\#}] = \left(\begin{array}{rr}
UH& UEH \\ U^{\#}EH & U^{\#}H \\
\end{array}\right).
\end{eqnarray*}
The columns of $[Z^{0*}Y\, (\Sigma Z^{0*})^{\#}Y^{\#}]$ are just a different set of $J$-orthonormal eigenvectors of $\mathcal{N}$, for $Y$ unitary. Notice that the matrix $H^{m \times r_{0*}}$  must have the structure $H=\bigl(\begin{smallmatrix} H_1^{r_{0*}\times r_{0*}} \\  \mathbf{0} \end{smallmatrix}\bigr)$, with $H_1$ being diagonal and full rank. Indeed, $2r_{0*} \leq \dim\, \Ker \,\mathcal{N} \leq \dim\, \Ker \, N \leq 2m \Rightarrow r_{0*} \leq m$, and $\Rank \, H = \Rank \, P_1 = \Rank \, P_2 = \frac{1}{2}\Rank \, P$, where $\Rank \, P = \min \, \{\Rank\, N, 2\,\Rank \, Z^{0*}\} = \min\, \{2n,2m,2r_{0*}\}=2r_{0*}$. Also, let $E_1$ be the $r_{0*}$-dimensional square diagonal matrix made up from the first $r_{0*}$ elements of the diagonal of $E$, and $U_1$ the $m \times r_{0*}$ matrix made up from the first $r_{0*}$ columns of $U$. We have then,
\begin{eqnarray*}
\left(\begin{array}{rr}
UH& UEH \\ U^{\#}EH & U^{\#}H \\
\end{array}\right) = \left(\begin{array}{rr}
U_1 H_1& U_1 E_1 H_1 \\ U_1^{\#}E_1 H_1 & U_1^{\#}H_1 \\
\end{array}\right) = \left(\begin{array}{ll}
U_1 & \mathbf{0} \\ \mathbf{0} & U_1^{\#} \\
\end{array}\right)
\left(\begin{array}{rr}
H_1 & E_1 H_1 \\ E_1 H_1 & H_1 \\
\end{array}\right),
\end{eqnarray*}
from which we conclude that
\begin{eqnarray}
N [Z^{0*}Y\, (\Sigma Z^{0*})^{\#}Y^{\#}] = \left(\begin{array}{ll}
U_1 & \mathbf{0} \\ \mathbf{0} & U_1^{\#} \\
\end{array}\right)
\left(\begin{array}{rr}
H_1 & E_1 H_1 \\ E_1 H_1 & H_1 \\
\end{array}\right). \label{Bogoliubov SVD, special degenerate case}
\end{eqnarray}
Equation (\ref{Bogoliubov SVD, special degenerate case}) is exactly the sought after decomposition of $N$. The columns of $\bigl(\begin{smallmatrix} U_1 & \mathbf{0} \\ \mathbf{0} & U_1^{\#} \end{smallmatrix}\bigr)$ provide a set of $2r_{0*}$ $J$-orthonormal vectors (though not a basis of $\Ima(P)$), and $\bar{N}_{0*}=\bigl(\begin{smallmatrix} H_1 & E_1 H_1 \\ E_1 H_1 & H_1 \end{smallmatrix}\bigr)$. The form of $\bar{N}_{0*}$ suggests that for a zero eigenvalue of $\mathcal{N}$ whose corresponding eigenvector is not in $\Ker N$, the implementation of its coupling matrix is by a cavity with a port whose passive and active coupling coefficients are equal in absolute value.

We demonstrate the result for this special case with an example from \cite[Section 8]{nurjampet09}.
\begin{example}\label{example degenerate case}
Consider the 1-mode, 3-input system with
\[ M=\left(\begin{array}{cc}
\Delta & 0 \\ 0 & \Delta \\
\end{array}\right), \
N_1=N_2=\left(\begin{array}{c}
\sqrt{\kappa_1} \\ \sqrt{\kappa_2} \\ \sqrt{\kappa_3} \\
 \end{array}\right), \ \mathrm{and}\ S=I_3. \]
We have that $\mathcal{N}=0_{2 \times 2}$, and $[Z^{0*}\, (\Sigma Z^{0*})^{\#}]=I_2$. Hence, $P=N$. However, we also have that $P P^{\flat}= N N^{\flat}=0_{6 \times 6}$. A SVD of $P_1=N_1$ is given by
\[ N_1 =  U \left(\begin{array}{c} \sqrt{\kappa} \\ 0 \\ 0 \\ \end{array}\right) \cdot 1, \]
where $\kappa=\kappa_1 +\kappa_2 +\kappa_3$, and $U=[u_1 \, u_2 \, u_3]$, with
\begin{eqnarray*}
u_1&=&\frac{1}{\sqrt{\kappa}} \left(\begin{array}{r}
\sqrt{\kappa_1} \\ \sqrt{\kappa_2} \\ \sqrt{\kappa_3} \\ \end{array}\right),\
u_2=\frac{1}{\sqrt{\kappa_1 +\kappa_2}} \left(\begin{array}{r}
-\sqrt{\kappa_2} \\ \sqrt{\kappa_1} \\ 0 \\ \end{array}\right),\ \mathrm{and} \\
u_3&=&\frac{1}{\sqrt{\kappa\,(\kappa_1 +\kappa_2)}} \left(\begin{array}{r}
\sqrt{\kappa_1\,\kappa_3} \\ \sqrt{\kappa_2\,\kappa_3} \\ -(\kappa_1 +\kappa_2) \\ \end{array}\right).
\end{eqnarray*}
Then, (\ref{Bogoliubov SVD, special degenerate case}) becomes $N=P=\bigl(\begin{smallmatrix} u_1 & 0_{3 \times 1} \\ 0_{3 \times 1} & u_1 \end{smallmatrix}\bigr)
\bigl(\begin{smallmatrix} \sqrt{\kappa} & \sqrt{\kappa} \\ \sqrt{\kappa} & \sqrt{\kappa} \end{smallmatrix}\bigr)$, which is obvious. Since there are no other eigenvectors of $\mathcal{N}$, the Bogoliubov matrices $V$ and $W$ in the statement of Theorem \ref{Bogoliubov SVD} are assembled as follows. First, $W=[Z^{0*}\, (\Sigma Z^{0*})^{\#}]=I_2$. To construct $V$, we must complete the $J$-orthonormal set $\left\{ \bigl(\begin{smallmatrix} u_1 \\ 0_{3 \times 1}\end{smallmatrix}\bigr), \bigl(\begin{smallmatrix} 0_{3 \times 1}\\ u_1 \end{smallmatrix}\bigr)\right\}$ into a $J$-orthonormal basis of $\mathbb{C}^6$. The easiest way to do this is to use the other two columns of $U$, and set $V=\bigl(\begin{smallmatrix} U & \mathbf{0} \\ \mathbf{0} & U^{\#} \end{smallmatrix}\bigr)$. Then, $N$ has the decomposition
\begin{eqnarray*}
N=\left(\begin{array}{ll} U & \mathbf{0} \\ \mathbf{0} & U^{\#} \\ \end{array}\right)\hat{N}, \ \mathrm{with} \ \hat{N}_1=\hat{N}_2=\left(\begin{array}{c} \sqrt{\kappa} \\ 0 \\ 0 \\ \end{array}\right).
\end{eqnarray*}
This decomposition could have been surmised directly from the SVD of $P_1$, since $N=P$, in this example. From the form of $\hat{N}$, we see that it can be implemented by a cavity with a port whose passive and active coupling coefficients are equal. The reduced system has the Hamiltonian matrix $\hat{M}=W^{\dag}M W =M$, and no feedback is necessary to create it. Figure \ref{Active_Network_3} provides a graphical representation of the proposed implementation of the transfer function for this example.
%-------------------
\begin{figure}[!h]
\begin{center}
\scalebox{.4}{\includegraphics{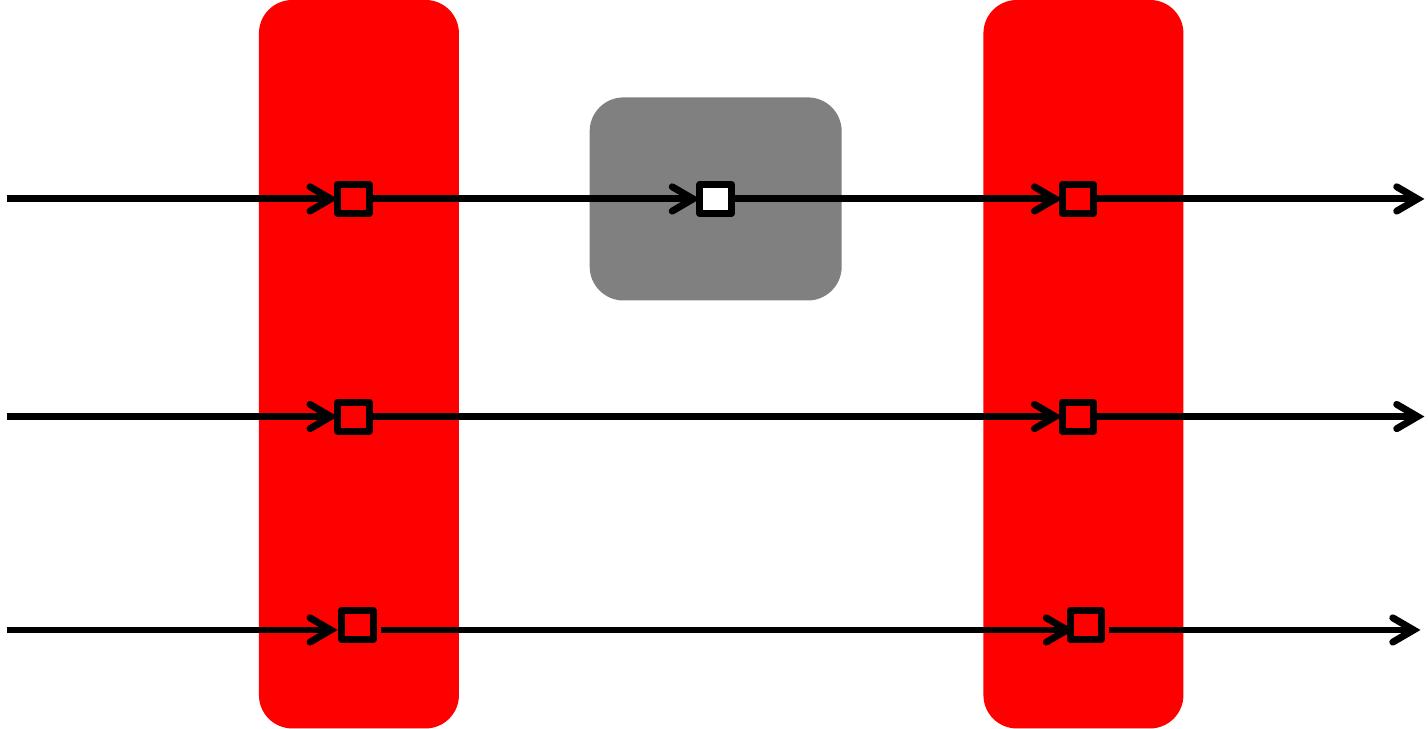}} \caption{Graphical representation of the proposed implementation of the transfer function of Example \ref{example degenerate case}.} \label{Active_Network_3}
\end{center}
\end{figure}
%-------------------
\end{example}

\bibliographystyle{ieeetr}
\bibliography{C:/Users/Symeon/Documents/AAA/Work/Latex/MyBibliographies/Linear_Quantum_Stochastic_Systems,C:/Users/Symeon/Documents/AAA/Work/Latex/MyBibliographies/Books,C:/Users/Symeon/Documents/AAA/Work/Latex/MyBibliographies/My_papers}
\end{document}